\documentclass[twoside,twocolumn,9pt]{article}
\usepackage{extsizes}
\usepackage[super,sort&compress,comma]{natbib} 
\usepackage[version=3]{mhchem}
\usepackage[left=1.5cm, right=1.5cm, top=1.785cm, bottom=2.0cm]{geometry}
\usepackage{balance}
\usepackage{mathptmx}
\usepackage{sectsty}
\usepackage{graphicx} 
\usepackage{lastpage}
\usepackage[format=plain,justification=justified,singlelinecheck=false,font={stretch=1.125,small,sf},labelfont=bf,labelsep=space]{caption}
\usepackage{float}
\usepackage{fancyhdr}
\usepackage{fnpos}
\usepackage[english]{babel}
\addto{\captionsenglish}{%
  
}
\usepackage{array}
\usepackage{droidsans}
\usepackage{charter}
\usepackage[T1]{fontenc}
\usepackage[usenames,dvipsnames]{xcolor}
\usepackage{setspace}
\usepackage[compact]{titlesec}
\usepackage{hyperref}
\usepackage{soul}

\usepackage{subcaption}

\definecolor{cream}{RGB}{222,217,201}

\begin{document}

\pagestyle{fancy}
\thispagestyle{plain}
\fancypagestyle{plain}{
\renewcommand{\headrulewidth}{0pt}
}

\makeFNbottom
\makeatletter
\renewcommand\LARGE{\@setfontsize\LARGE{15pt}{17}}
\renewcommand\Large{\@setfontsize\Large{12pt}{14}}
\renewcommand\large{\@setfontsize\large{10pt}{12}}
\renewcommand\footnotesize{\@setfontsize\footnotesize{7pt}{10}}
\makeatother

\renewcommand{\thefootnote}{\fnsymbol{footnote}}
\renewcommand\footnoterule{\vspace*{1pt}%
\color{cream}\hrule width 3.5in height 0.4pt \color{black}\vspace*{5pt}} 
\setcounter{secnumdepth}{5}

\makeatletter 
\renewcommand\@biblabel[1]{#1}            
\renewcommand\@makefntext[1]%
{\noindent\makebox[0pt][r]{\@thefnmark\,}#1}
\makeatother 
\renewcommand{\figurename}{\small{Fig.}~}
\sectionfont{\sffamily\Large}
\subsectionfont{\normalsize}
\subsubsectionfont{\bf}
\setstretch{1.125} 
\setlength{\skip\footins}{0.8cm}
\setlength{\footnotesep}{0.25cm}
\setlength{\jot}{10pt}
\titlespacing*{\section}{0pt}{4pt}{4pt}
\titlespacing*{\subsection}{0pt}{15pt}{1pt}

\fancyfoot{}
\fancyfoot[RO]{\footnotesize{\sffamily{1--\pageref{LastPage} ~\textbar  \hspace{2pt}\thepage}}}
\fancyfoot[LE]{\footnotesize{\sffamily{\thepage~\textbar\hspace{4.65cm} 1--\pageref{LastPage}}}}
\fancyhead{}
\renewcommand{\headrulewidth}{0pt} 
\renewcommand{\footrulewidth}{0pt}
\setlength{\arrayrulewidth}{1pt}
\setlength{\columnsep}{6.5mm}
\setlength\bibsep{1pt}

\makeatletter 
\newlength{\figrulesep} 
\setlength{\figrulesep}{0.5\textfloatsep} 

\newcommand{\topfigrule}{\vspace*{-1pt}%
\noindent{\color{cream}\rule[-\figrulesep]{\columnwidth}{1.5pt}} }

\newcommand{\botfigrule}{\vspace*{-2pt}%
\noindent{\color{cream}\rule[\figrulesep]{\columnwidth}{1.5pt}} }

\newcommand{\dblfigrule}{\vspace*{-1pt}%
\noindent{\color{cream}\rule[-\figrulesep]{\textwidth}{1.5pt}} }

\makeatother

\twocolumn[
  \begin{@twocolumnfalse}
\vspace{1em}
\sffamily
\begin{tabular}{m{4.5cm} p{13.5cm} }

& \noindent\LARGE{\textbf{Probing Disorder in 2CzPN using Core and Valence States}} \\ 
\vspace{0.3cm} & \vspace{0.3cm} \\

 & \noindent\large{Nathalie~K.~Fernando,\textit{$^{a}$} Martina~Stella,\textit{$^{bc}$} William Dawson,\textit{$^{d}$}, Takahito Nakajima,\textit{$^{d}$} Luigi~Genovese,\textit{$^{e}$} Anna~Regoutz,\textit{$^{a}$}  and Laura~E.~Ratcliff\textit{$^{bf\ast}$}} \\
 
& \noindent\normalsize{Molecules which exhibit thermally activated delayed fluorescence (TADF) show great promise for use in efficient, environmentally-friendly OLEDs, and thus the design of new TADF emitters is an active area of research.  However, when used in devices, they are typically in the form of disordered thin films, where both the external molecular environment and thermally-induced internal variations in parameters such as the torsion angle can strongly influence their electronic structure. In this work, we use density functional theory and X-ray photoelectron spectroscopy to investigate the impact of disorder on both core and valence states in the TADF emitter 2CzPN. By simulating gas phase molecules displaying varying levels of disorder, we assess the relative sensitivity of the different states to factors such as varying torsion angle. The theoretical results for both core and valence states show good agreement with experiment, thereby also highlighting the advantages of our approach for interpreting experimental spectra of large aromatic molecules, which are too complex to interpret based solely on experimental data.} \\
\end{tabular}

 \end{@twocolumnfalse} \vspace{0.6cm}

  ]

\renewcommand*\rmdefault{bch}\normalfont\upshape
\rmfamily
\section*{}
\vspace{-1cm}


\footnotetext{\textit{$^{a}$~Department of Chemistry, University College London, 20 Gordon Street, London, WC1H 0AJ, United Kingdom.}}
\footnotetext{\textit{$^{c}$~The Abdus Salam International Centre for Theoretical Physics, Condensed Matter and Statistical Physics, 34151 Trieste, Italy.}}
\footnotetext{\textit{$^{b}$~Department of Materials, Imperial College London, London SW7 2AZ, United Kingdom.}}
\footnotetext{\textit{$^{d}$~RIKEN Center for Computational Science, Kobe, Japan.}}
\footnotetext{\textit{$^{e}$~Univ.\ Grenoble Alpes, CEA, IRIG-MEM-L\_Sim, 38000 Grenoble, France.}}
\footnotetext{\textit{$^{f}$~Centre for Computational Chemistry,
School of Chemistry, University of Bristol, Bristol BS8 1TS,
United Kingdom. Email: laura.ratcliff@bristol.ac.uk}}




\section{Introduction}

Organic light-emitting diodes (OLEDs) have established themselves as a leading technology in a variety of commercial devices, ranging from lighting to full-color displays\cite{DeAngelis2014,Tao2014,Wong2017}. However, their internal quantum efficiency (IQE) is not always optimal due to the unfavourable spin statistics observed when charges combine to create excitons~\cite{Hiroki2012}, a situation which is typically remedied by employing emitters containing heavy metals such as Ir or Pt. A promising alternative to such potentially costly and environmentally harmful compounds is provided by a novel class of OLEDs which can be fully organic, and rely on the mechanism of thermally activated delayed fluorescence (TADF). TADF emission is characterised by reverse intersystem crossing taking place between non-radiative triplet states, which act as a reservoir to the first excited radiative singlet state, giving a maximum potential IQE of 100\%~\cite{Zhang2015}. Such a mechanism is facilitated by a small singlet-triplet energy splitting, $\Delta E_{\mathrm{ST}}$. Being able to predict which molecules display this mechanism, as well as understand which factors influence $\Delta E_{\mathrm{ST}}$, is therefore crucial for determining the structural features that facilitate TADF. 

Established computational tools such as density functional theory (DFT)~\cite{Hohenberg1964,Kohn1965} are commonly used to model the excited states of ideal isolated gas phase TADF molecules. However, these molecules interact with their neighbouring molecules in the thin films used in devices, and are also susceptible to conformational fluctuations due to thermal effects. Therefore, in order to gain a more realistic description of TADF materials it is crucial to also properly account for thermal fluctuations of the molecular conformations and microscopic electronic polarization effects in amorphous materials~\cite{Desilva2019, Kim2021}. For example, Olivier \emph{et al.}~\cite{Olivier2017} combined classical molecular dynamics (MD) with time-dependent DFT (TD-DFT) to study two TADF molecules, including 2CzPN (1,2-bis(carbazol-9-yl)-4,5-dicyanobenzene). They demonstrated the influence of conformational fluctuations on emission properties, highlighting the effectiveness of coupling experimental observations with theoretical insights for investigating disorder in TADF materials. 

While understanding the effects of disorder on excited states is crucial for applications, excited state methods are challenging in terms of both cost and accuracy. Nonetheless, meaningful information can also be gained by looking at the influence of disorder on the ground state electronic structure, an area which is relatively unexplored for many TADF emitters.  To this end, we present a combined computational and experimental investigation into the ground state electronic structure of the prototypical TADF emitter 2CzPN, including both core and valence states. While valence states can be used to gain useful insights into the effects of disorder, core states have the additional advantage that they can be directly mapped to specific atomic environments, and thus experimental approaches such as core X-ray photoelectron spectroscopy (XPS) can be a useful tool for acting as a local probe of disorder. However, the interpretation of core XPS typically relies on comparison to reference spectra, which is challenging in the case of large molecules containing many different local chemical environments.  Despite having received relatively little attention compared to valence state calculations, the theoretical prediction of core binding energies (BEs) is invaluable for the interpretation of core XPS.  While calculations based on more expensive quantum chemical approaches or methods such as GW~\cite{Golze2020} show great promise, DFT has the advantage of being less computationally expensive, thereby allowing the treatment of much larger systems.

One promising DFT-based method for core BE calculations is $\Delta$SCF, in which an electron is removed from a core state of interest, with the corresponding BE being defined as the difference between the energy of the core-excited state and that of the ground state. It thereby enables final state effects to be taken into account with modest computational cost. $\Delta$SCF has been successfully employed for benchmark calculations of a range of molecules, however, it is not yet widely used in applications, with most calculations also having been restricted to small molecules.  One challenge for $\Delta$SCF in Gaussian basis codes is the need to either use very large or specifically constructed basis sets which are adapted for core-excited calculations, e.g.\ by including basis functions for the $Z+1$ element~\cite{Hanson-Heine2018}. We recently introduced an alternative approach, which uses an adaptive multi-wavelet basis, as implemented in the MADNESS code~\cite{Harrison2016}.  The multi-wavelet basis leads to a guaranteed, high precision without requiring any user intervention or specialist knowledge.  In combination with a mixed all-electron/pseudopotential scheme, this facilitates a highly accurate and computationally efficient approach~\cite{Pi2020}, which also avoids the common problem of core-hole hopping and associated convergence problems.  We have previously applied this approach to both simple~\cite{Pi2020} and aromatic amino acids~\cite{Regoutz2020}, with final state effects playing a significant role with the latter, so that experimental results cannot be interpreted using chemical intuition alone.  In this work we investigate the effects of disorder on the core BEs of 2CzPN, demonstrating how our approach may be combined with experiment to provide a powerful tool for investigating core states in large molecules.

\section{Methods}

\subsection{Atomic Structure}

\begin{figure}[!ht]
\centering
\begin{subfigure}[t]{0.45\linewidth}
\centering
\includegraphics[width=0.75\textwidth]{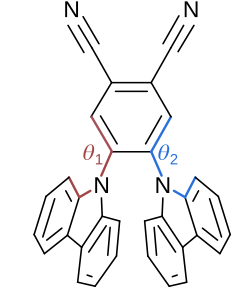}
\caption{Atomic structure and definition of torsion angles, $\theta_1$ and $\theta_2$.}
\label{fig:angles}
\end{subfigure}
\hspace{6pt}
\begin{subfigure}[t]{0.45\linewidth}
\centering
\includegraphics[width=0.95\textwidth]{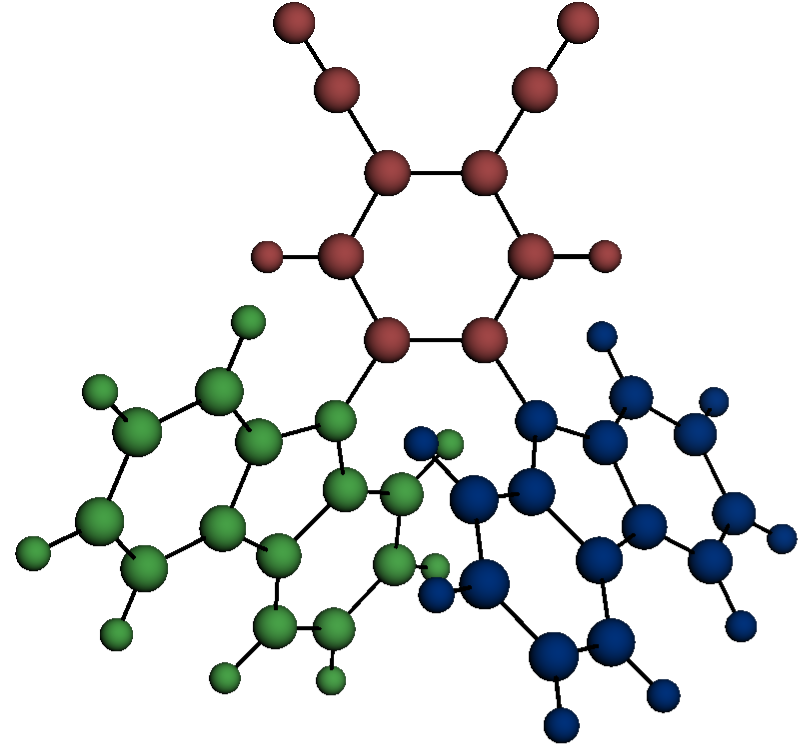}
\caption{Visualisation of the three subfragments in the relaxed  molecule.}
\label{fig:subfrags}
\end{subfigure}
\begin{subfigure}[t]{0.45\linewidth}
\centering
\includegraphics[width=0.75\textwidth]{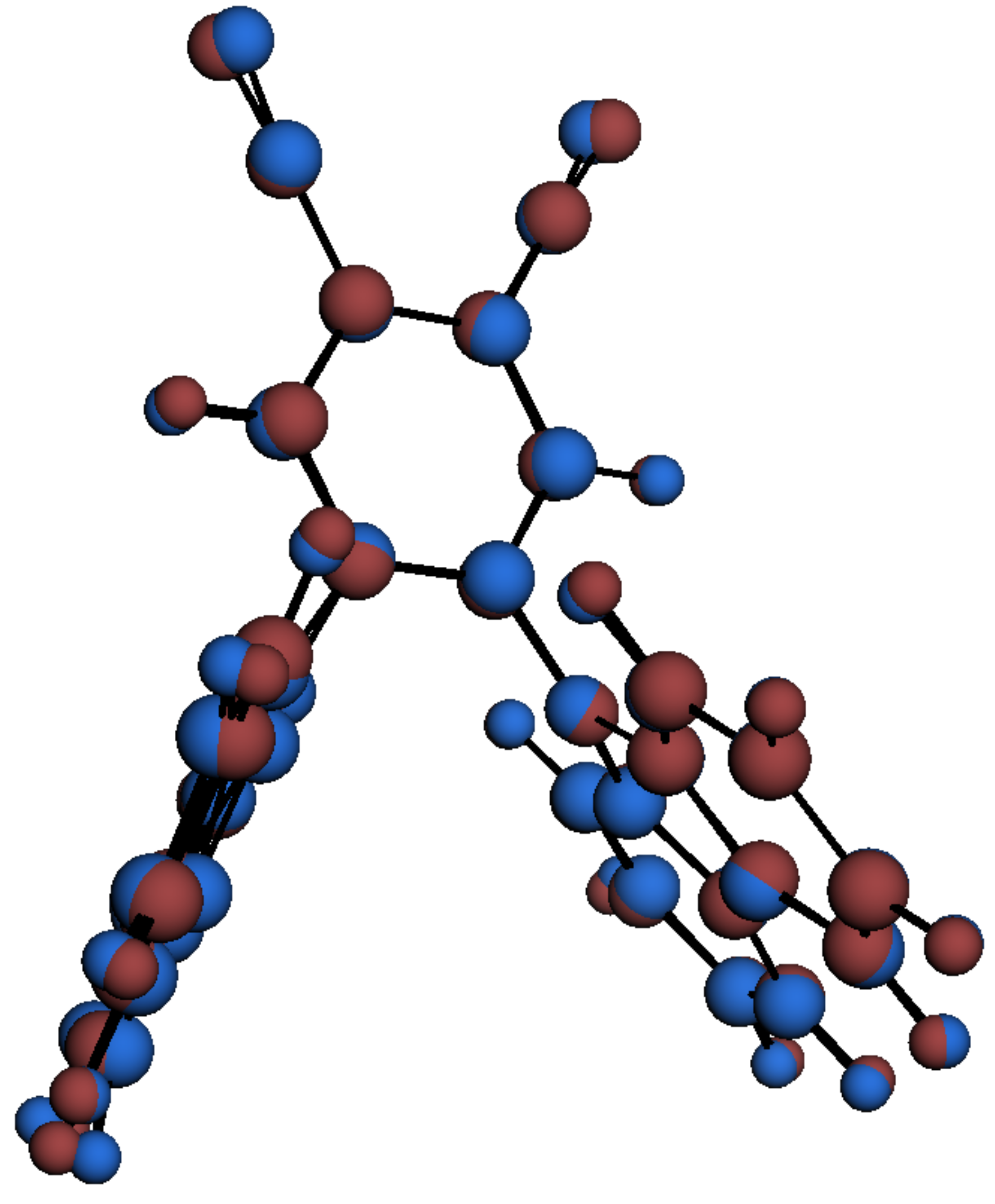}
\caption{Rigidification example, where blue (red) atoms indicate the original (rigidified) structure.}
\label{fig:rigidification}
\end{subfigure}
\hspace{6pt}
\begin{subfigure}[t]{0.45\linewidth}
\centering
\includegraphics[width=0.95\textwidth]{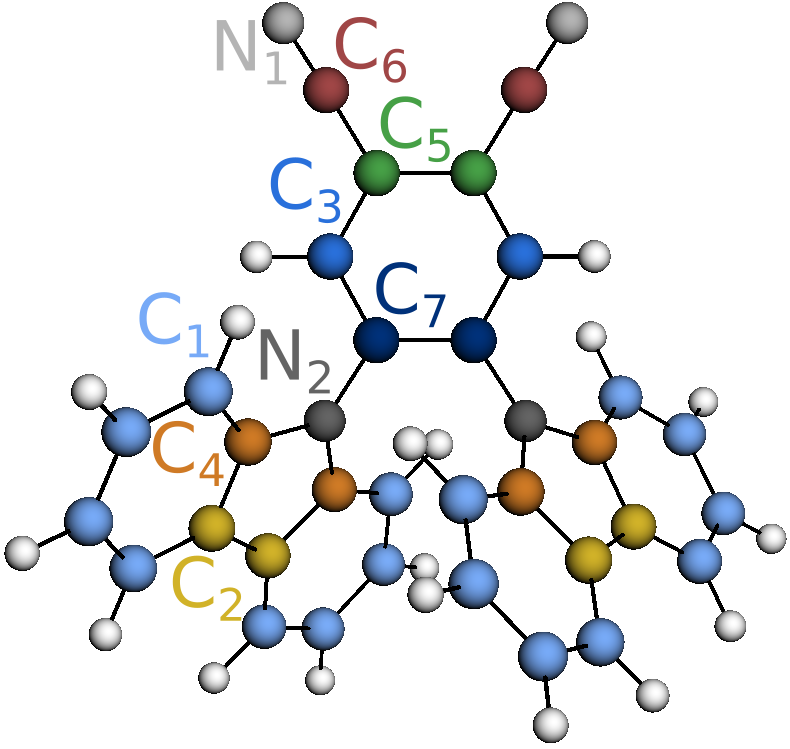}
\caption{Visualisation of the relaxed molecule, where the colours denote similar core binding energies.}
\label{fig:core_colours}
\end{subfigure}
\caption{Depiction of the atomic structure of 2CzPN, showing the torsion angles in (a). The subfragments depicted in (b) are used to perform the rigidification in (c). (d) shows the atoms grouped according to similar chemical environments and thus core binding energies, and their corresponding labels. \label{fig:structure}}
\end{figure}

The atomic structure of 2CzPN is depicted in Fig.~\ref{fig:angles}, and is characterised by three `subfragments' (highlighted in Fig.~\ref{fig:subfrags}): a central dicyano-subsituted phenylene core, which is linked to two carbazole units, whose relative orientations are described by torsion angles $\theta_1$ and $\theta_2$.  The energy barrier is small enough such that thermal effects may lead to disorder in the form of diverse torsion angles being present in thin films of 2CzPN.  This variation can in turn influence both the ground and excited state electronic structure.  Starting from an amorphous MD snapshot extracted from an equilibrated simulation, which was previously used to explore such effects on excited states~\cite{Olivier2017}, in the following we assess the influence of various levels of disorder on the ground state electronic structure of 2CzPN.  In order to reduce the computational cost, while still sampling enough of the conformational space, a cluster of 500 molecules was extracted out of a given 1000 molecule MD snapshot.  The focus of this work is the internal disorder present in 2CzPN, and thus each of the 500 molecules were calculated in separate gas phase calculations without further relaxation, while also being compared with the relaxed gas phase molecule.  Free boundary conditions are used throughout.

Since both torsion angles may vary independently, using the angle alone does not provide a single parameter which can characterise the structure of a molecule.  Although one could use the average value, some information is lost when comparing how similar two structures are.  We therefore propose an alternative parameter, namely the cost function $J$, which is defined as
\begin{equation}\label{eq:wahba}
J_{\alpha\beta}=\frac{1}{N_{\mathrm{at}}}\sum_{a=1}^{N_{\mathrm{at}}}||\mathbf{R}_a^\alpha-\sum_{b=1}^{N_{\mathrm{at}}}\mathcal{R}^{\beta\rightarrow \alpha}_{ab}\mathbf{R}_a^\beta||^2\;,
\end{equation}
where $\mathcal{R}^{\beta\rightarrow \alpha}_{ab}$ is a rotation matrix between two instances of a molecule characterised by their atomic coordinates $\mathbf{R}^{\alpha}$ and $\mathbf{R}^{\beta}$, and $N_{\mathrm{at}}$ is the number of atoms in the molecule.  The optimal rotation, $\mathcal{R}^{\beta\rightarrow \alpha}_{ab}$, may be efficiently found by minimising $J$~\cite{Wahba1965,Kabsch1978,Markley1988}, where $J=0$ implies a rigid transformation.  $J$ has previously been employed in BigDFT as a measure of similarity between fragment geometries for the purposes of reusing basis functions between calculations~\cite{Ratcliff2015a,Ratcliff2019b,Ratcliff2020}, and thus provides a computationally inexpensive means of comparing extracted molecules. At present, inversion symmetry is not taken into account, this could, however, be included in the future.

A small value of $J_{\alpha\beta}$ implies that molecules $\alpha$ and $\beta$ have similar torsion angles, while also not showing any other significant differences in geometry.  Indeed, beyond variations in the torsion angles, thermal fluctuations also result in other distortions, which may be captured by the use of $J$. This is evidenced by the fact that in the relaxed molecule, the three subfragments shown in Fig.~\ref{fig:subfrags} are close to planar, while the equivalent subfragments in the molecules extracted from the MD snapshot show noticeable distortions, and are no longer planar. In order to assess the impact of this additional disorder, for each of the 500 extracted molecules, we also generate a `rigidified' molecule.  This is achieved using the relaxed molecule as a template.  First, the rototranslations between each of the subfragments in the template and those of the target molecule are identified.  Then the rototranslations are applied to each subfragment in the template molecule, thereby constructing a new molecule which preserves the internal structure of the subfragments of the relaxed molecule. An example molecule to which this process has been applied is depicted in Fig.~\ref{fig:rigidification}.

Following the rigidification, there are thus two sets of unaltered and rigidified molecules, containing 500 molecules each.  In order to reduce the number of computationally intensive calculations to be performed, while nonetheless maintaining a balanced coverage of configuration space, the cost function may be used to identify a subset of distinct molecules for each set.  This is achieved by means of a hierarchical clustering approach, as implemented in SciPy~\cite{Virtanen2020}, using the matrix $J_{\alpha\beta}$ as the linkage matrix, where $\alpha$ and $\beta$ run over all molecules in the set. Based on a threshold value of $J$, the molecules are grouped into clusters, with a lower value of $J$ giving rise to more clusters. One molecule from each cluster is then selected as a representative. In this work, the first molecule from each cluster is selected, but alternative choices could be used in future.  The representative molecules are then used to perform both the core calculations and hybrid functional valence state calculations, where the computational cost of calculating all molecules would be very high. The clustering is performed for the unmodified extracted molecules only; the corresponding rigidified molecules are directly employed.

Finally, as an independent comparison to the extracted molecules, a set of explicitly rotated molecules were also generated.  Initially these were generated from the DFT-relaxed molecule, keeping the central subfragment fixed, and rotating the two carbazole units in tandem.  However, the interaction between the two carbazole units in the relaxed molecule is such that a small degree of asymmetry is introduced, resulting in an asymmetric potential energy surface (see Fig.~S3 in the Supplementary Information). Therefore, the molecule was also relaxed with constraints on the symmetry and torsion angle, with $\theta_1 \equiv \theta_2 \equiv 90^\circ$. Twenty additional rotated molecules, with angles ranging from $40^\circ$ to $140^\circ$ were then generated starting from the constrained molecule, keeping $\theta_1\equiv\theta_2$. Due to the imposed symmetry, only molecules with $\theta_{\mathrm{Av.}}\leq 90^{\circ}$ were explicitly calculated. 
The symmetry-constrained molecule was also used as an alternative template for performing the rigidification, however the differences were found to be small (less than 0.1~eV average difference in the PBE-calculated frontier orbital energies), and so the unconstrained relaxed molecule was retained as the template for the rigidification.

To summarise, three sets of molecules with varying levels of disorder were therefore generated:
\begin{enumerate}
    \item Rotated molecules: the torsion angle may vary, but $\theta_1\equiv\theta_2$, while the subfragments are kept rigid;
    \item Rigid molecules: both torsion angles may vary independently, while rigidity is enforced in the subfragments;
    \item Extracted molecules: no constraints are present, with all atoms able to move freely.
\end{enumerate}
By calculating core and valence quantities for all three sets of molecules, we are therefore able to assess the relative importance of the different types of disorder present in 2CzPN.

\subsection{Density Functional Theory}\label{sec:dft}

Valence state calculations were performed using the BigDFT code~\cite{Ratcliff2020}, which uses a Daubechies wavelet basis set~\cite{Daubechies1992}. Calculations employed the cubic scaling version of BigDFT~\cite{Genovese2008}, using HGH-GTH pseudopotentials (PSPs)~\cite{Goedecker1996,Hartwigsen1998} with non-linear core corrections~\cite{Willand2013}. A wavelet grid spacing of 0.5~bohr was used, with coarse (fine) multipliers of 5 (7). The PBE functional was used for all molecules~\cite{Perdew1996}, while only the representative molecules described above were calculated using the hybrid PBE0 functional~\cite{Adamo1999}. The molecule was relaxed using a smaller wavelet grid spacing of 0.45~bohr, using a maximum force threshold of 0.02~eV/\AA. Projected densities of states (PDOS) were calculated using a Mulliken-style population analysis, with Gaussian smearing of 0.44~eV applied to reflect the experimental broadening.  For comparison with the valence XPS measurements, Scofield photoionisation cross sections for 1.48667~keV~\cite{Scofield1973, Kalha2020} were applied to the calculated PDOS using the Galore package~\cite{Jackson2018}. Theoretical spectra were aligned and normalised to the lowest binding energy peak in the experimental spectrum.

Core binding energies (BEs) were calculated using the MADNESS code~\cite{Harrison2016}, which uses a multi-wavelet approach to achieve guaranteed high precision and can be used for all-electron (AE), PSP, or mixed AE/PSP calculations~\cite{Ratcliff2019a}. Calculations were performed at both the Koopmans' and $\Delta$SCF level of theory, using a recently developed approach in which the core-excited atom is treated at the AE level, and the remainder of the atoms are treated at the PSP level, as described in detail in Ref.~\cite{Pi2020}. This approach allows explicit access to the core states of interest, while forcing the core hole to remain localised, and at the same time reducing the computational cost compared to a pure AE approach. Following refs~\cite{Pi2020,Regoutz2020}, ground state calculations used a wavelet threshold of 10$^{-4}$ followed by $10^{-6}$ (wavelet orders $k=6$ and $k=8$ respectively), with core hole calculations using a wavelet threshold of $10^{-6}$ ($k=8$). HGH-GTH PSPs were used, while all calculations were spin restricted. Except where otherwise stated, calculations employed PBE. Since we are only interested in relative BEs, relativistic effects were neglected. All other parameters are the same as in refs~\cite{Pi2020,Regoutz2020}. A combination of 0.44~eV Gaussian smearing based on the experimental resolution and a 0.2~eV Lorentzian smearing to reflect lifetime contributions was applied to the calculated BEs. C~$1s$ theoretical spectra were aligned and normalised to the main experimental peak, while N~$1s$ spectra were aligned to the higher BE peak.

\subsection{Experimental Approach}

X-ray photoelectron spectra were collected on a Thermo Scientific K-Alpha+ spectrometer at Imperial College London, London, UK. The system uses a microfocused Al K$\alpha$ X-ray source with a photon energy of 1486.7~eV. The X-ray source is operated at 12~kV and 6~mA, with a base pressure in the analysis chamber of 2$\times10^{-9}$~mbar. The X-ray spot size used for the measurements was 400~$\mu$m. The 2CzPN powder, acquired from Ossila (Sublimed \textgreater99\% (HPLC)), was mounted on the sample plate using conductive carbon tape and the spectrometer flood gun was used to compensate for possible sample charging. 

In order to avoid any beam induced sample damage, the sample was rastered and the main core level spectra (C~$1s$ and N~$1s$) and valence spectra measured at equidistant spots across a defined rectangular region on the sample ($6\times3$ for the core and $5\times3$ for the valence spectra). This method enabled each spot to be irradiated for a significantly shorter time than would be necessary in a single spot measurement, achieving equally good signal to noise. The core level spectra were collected at a pass energy of 20~eV and a dwell time of 50~ms at each measurement point, whilst the valence spectra were measured at 40~eV pass energy and 75~ms dwell time. The survey spectrum, obtained from a single spot, was measured at 200~eV pass energy. The \emph{Multiplex} function on the data acquisition software (Thermo Scientific Avantage) enabled the measurement of single spectra of each core level or valence band until the predefined total number of scans had been collected, in order to equally distribute the X-ray dose received.

\section{Results}

\subsection{Atomic Structure}\label{sec:structure}

\begin{figure}[!ht]
\centering
\includegraphics[scale=0.37]{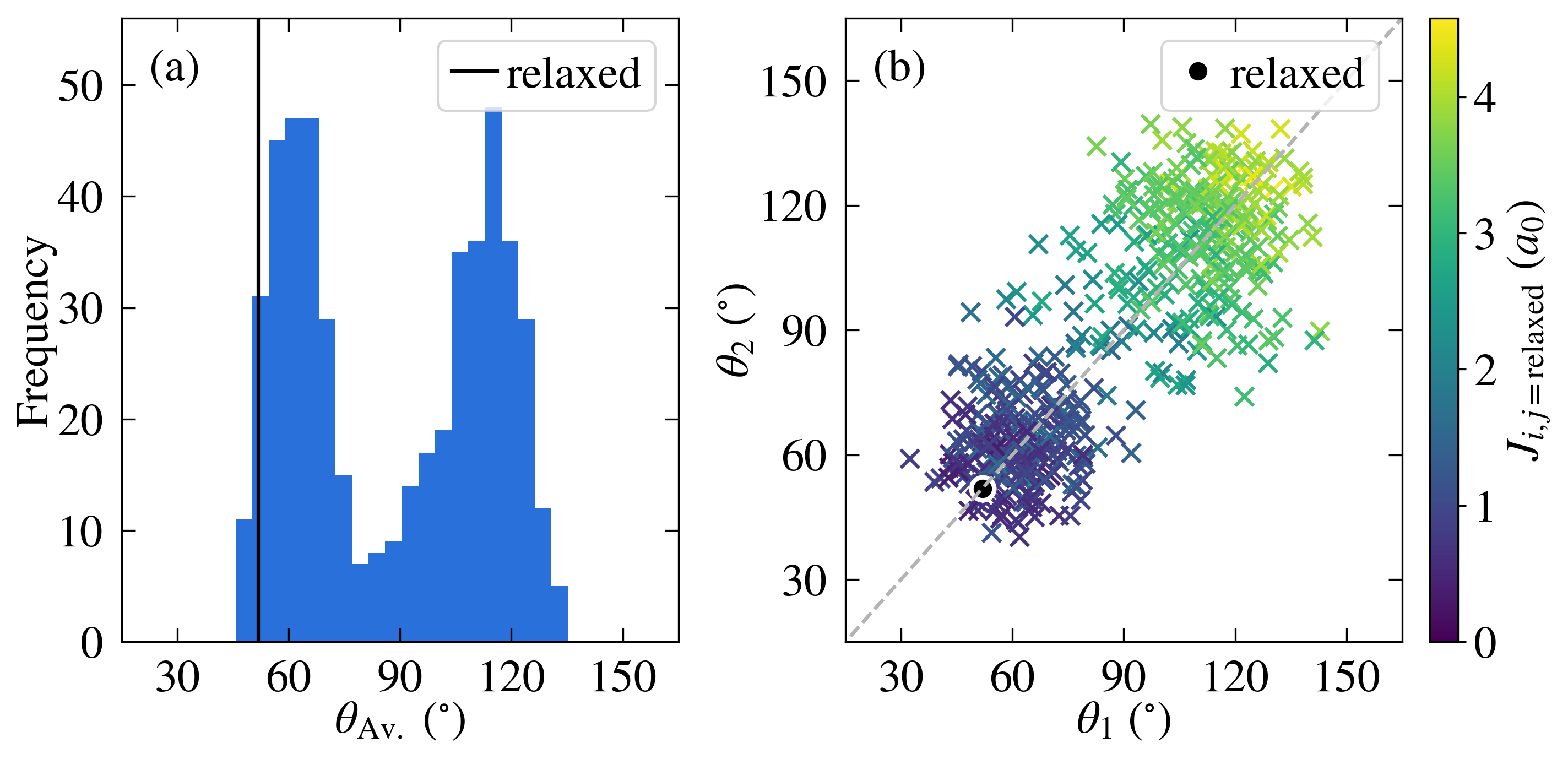}
\caption{(a) Distribution of average torsion angle, $\theta_{\mathrm{Av.}}$, across the 500 unmodified extracted molecules; and (b) comparison between the two torsion angles, $\theta_1$ and $\theta_2$, with the corresponding cost function $J_{i,j=\mathrm{relaxed}}$ between each molecule and the relaxed gas phase structure. Also depicted on both plots are the corresponding angles for the relaxed molecule. \label{fig:angle_plot}}
\end{figure}

Before discussing the electronic structure of 2CzPN, we first consider the atomic structure, including the variation in torsion angle. Although fewer molecules are considered in this work compared to the full trajectory of Ref.~\cite{Olivier2017}, as shown in Fig.~\ref{fig:angle_plot}a, 500 molecules are nonetheless sufficient to see the observed bimodal distribution in the average torsion angle, with peaks at around $60^{\circ}$ and $120~^{\circ}$.  In the PBE-relaxed molecule $\theta_1=\theta_2=51.9^{\circ}$; although there has been shown to be some sensitivity with respect to functional choice, this value nonetheless falls within the range of values calculated using different functionals~\cite{Lin2017}. In contrast to the relaxed molecule, Fig.~\ref{fig:angle_plot}b shows significant differences between $\theta_1$ and $\theta_2$ for a large number of the molecules, by more than $50~^{\circ}$ in some cases. At the same time, there is also disorder present in the bond lengths, with C-N and C-C bonds showing a wider distribution of ranges compared to the relaxed molecule (see Fig.~S1 in the Supplementary Information), further supporting the need for a metric for comparing molecules which goes beyond the average torsion angle.  To this end, Fig.~\ref{fig:angle_plot}b also shows the relationship between the torsion angles and the cost function $J$ between each molecule and the relaxed structure. As expected, molecules with smaller torsion angles, close to the relaxed value, have a smaller cost function value, in other words $J$ very loosely follows the difference in average torsion angle, while also taking into account other differences in structure.

\begin{figure*}[!ht]
\centering
\hspace{6pt}
\begin{subfigure}[t]{0.24\linewidth}
\includegraphics[scale=0.37]{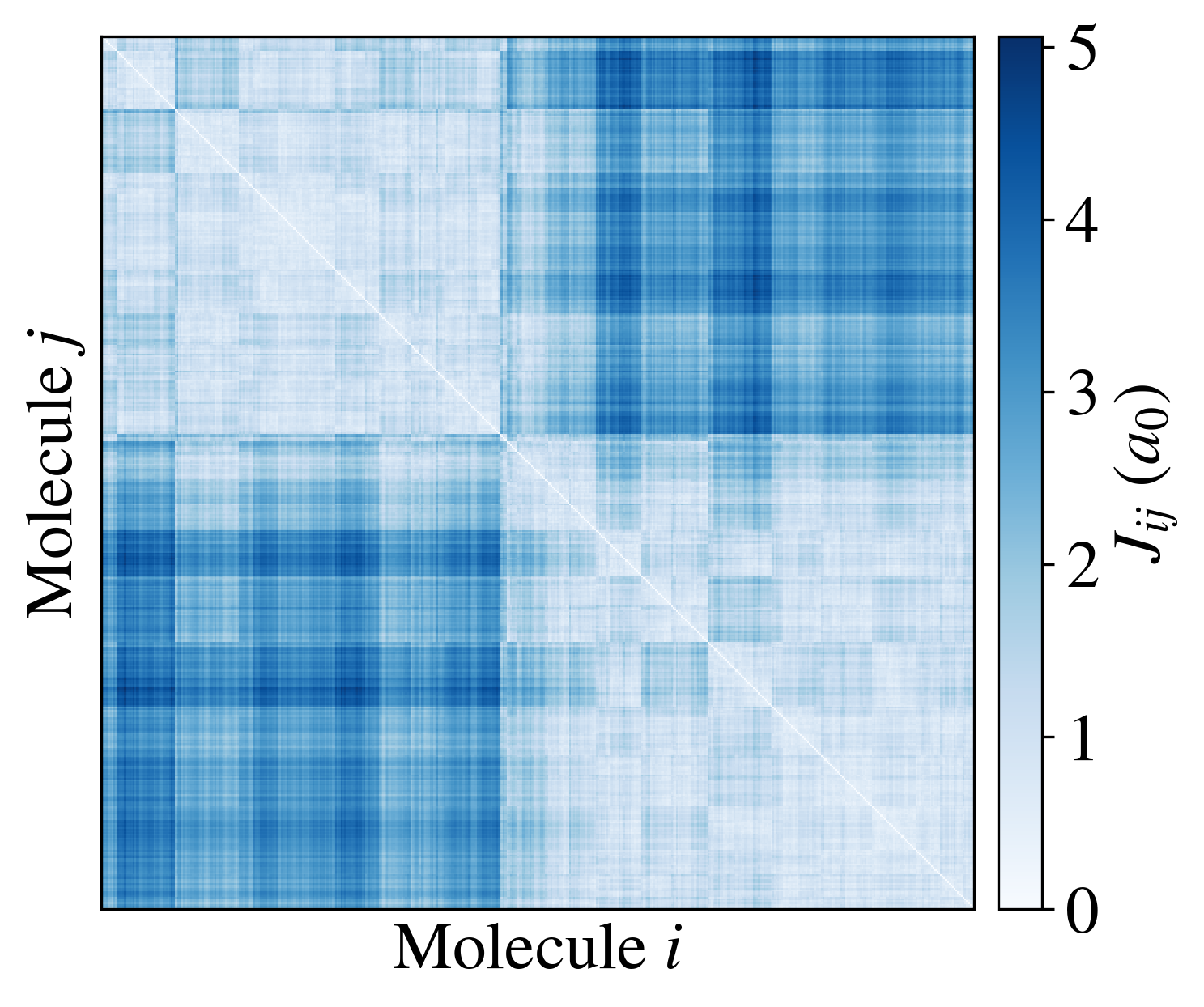}
\caption{Cost function matrix, $J_{ij}$, between molecules $i$ and $j$.}
\label{fig:jmat}
\end{subfigure}
\hspace{2pt}
\begin{subfigure}[t]{0.72\linewidth}
\includegraphics[scale=0.37]{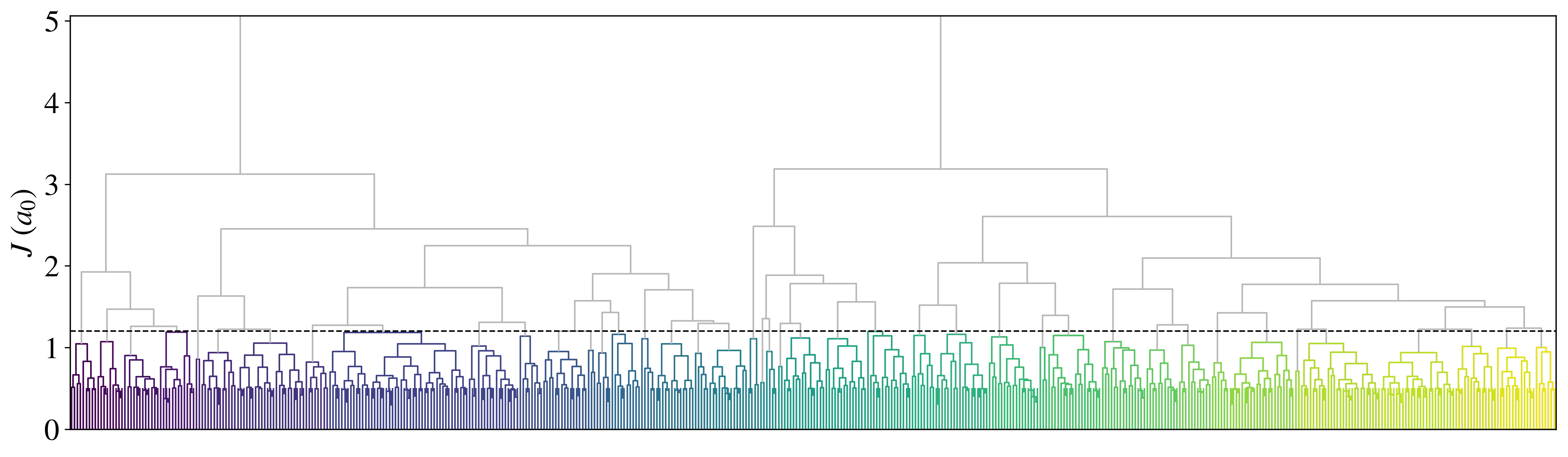}
\caption{Dendrogram showing the hierarchical relationship between clusters of similar molecules, where different colours denote a different cluster. The dashed black line denotes the threshold value of $J$ used to perform the clustering.}
\label{fig:dendrogram}
\end{subfigure}
\vspace{12pt}
\begin{subfigure}[t]{1.0\linewidth}
\includegraphics[width=1.0\textwidth]{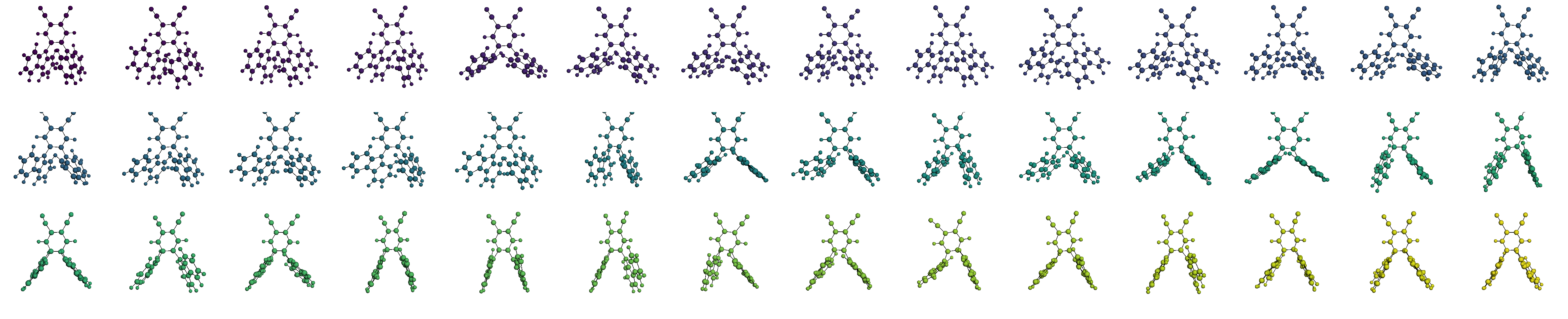}
\caption{Depiction of the first molecule from each of the clusters, where the colours correspond to those in (b).}
\label{fig:molecules}
\end{subfigure}
\caption{Depiction of (a) the cost function matrix, (b) the resulting hierarchical relationship, and (c) the representative molecules. \label{fig:sorting}}
\end{figure*}

The full cost function matrix between the 500 extracted molecules is depicted in Fig.~\ref{fig:jmat}, with the corresponding dendrogram showing the hierarchical clusters in Fig.~\ref{fig:dendrogram}. By varying the cost function threshold, the number of clusters can also be varied.  Using a cost function threshold of $J=1.2$~$a_0$\ results in 42 clusters of similar molecules; the first molecule of each cluster is depicted in Fig.~\ref{fig:molecules}. These molecules show a relatively well distributed range of torsion angles, while there is also a mix of molecules with both similar torsion angles and larger differences between the two torsion angles (see also Fig.~S2 in the Supplementary Information).  Similarly, the molecules are well distributed across the total energy landscape, with sufficient sampling to give a reasonable approximation of both the probability density and average energy for both the unmodified extracted and rigidified molecules (Fig.~S4 in the Supplementary Information).  This gives confidence that the chosen threshold is sufficient to capture the variety of torsion angles and other disorder present in the full dataset, an assumption which is further validated in the following investigation of the valence states.

\subsection{Valence States}\label{sec:valence}

Before investigating the core states, we first consider the valence states as well as the band gap. The relatively lower computational cost of calculating valence states allows the choice of selecting molecules to be validated, while it is also interesting to compare the relative sensitivity to disorder between core and valence states.

\subsubsection{Frontier Orbitals}\label{sec:frontier_orbitals}

\begin{figure*}[!ht]
\centering
\includegraphics[scale=0.37]{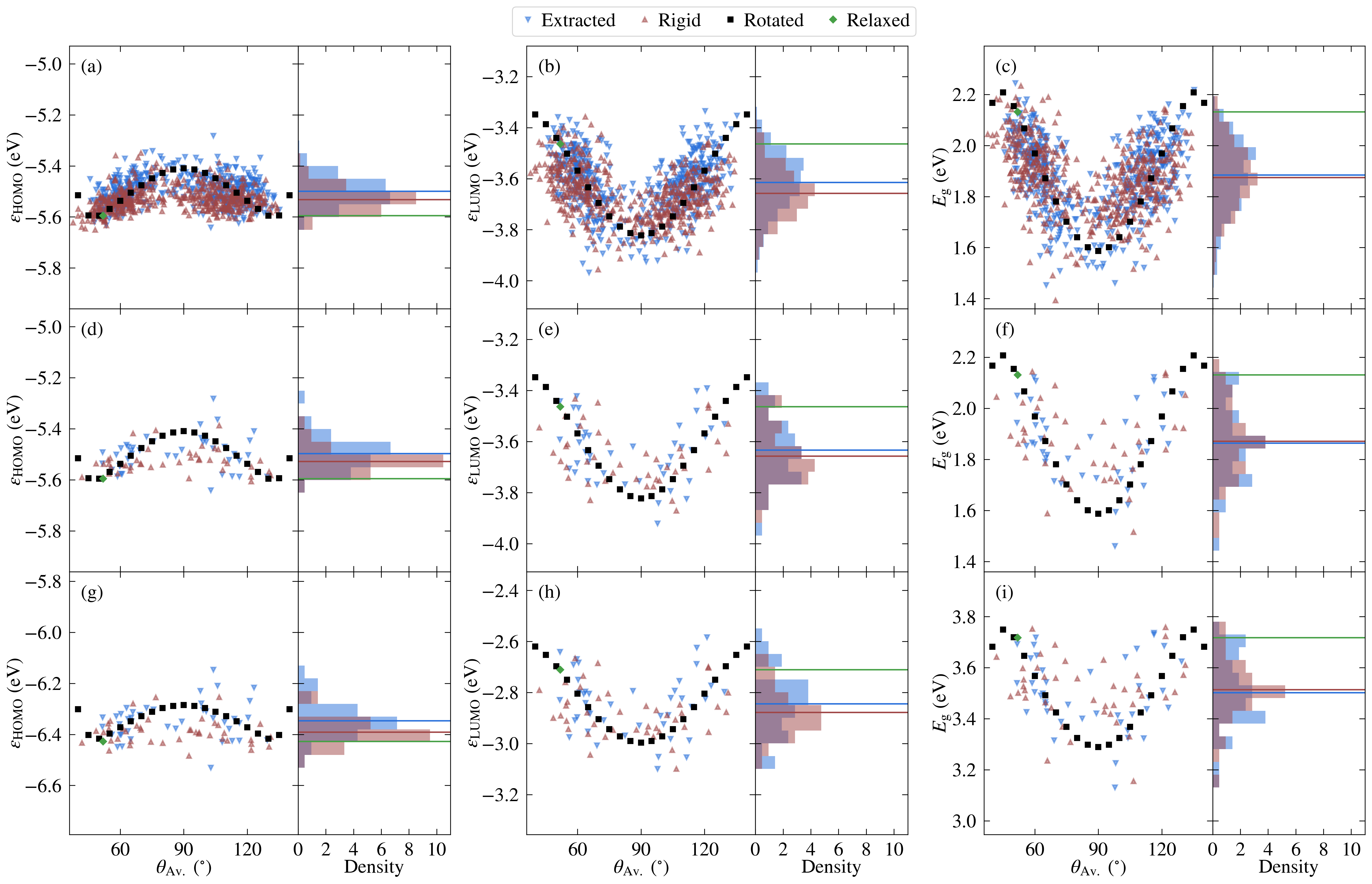}
\caption{Variation in the highest occupied molecular orbital energy, $\varepsilon_{\mathrm{HOMO}}$ [left], the lowest unoccupied molecular orbital energy, $\varepsilon_{\mathrm{LUMO}}$ [centre], and the band gap, $E_{\mathrm{g}}$ [right], with respect to the average torsion angle $\theta_{\mathrm{Av.}}$, as well as the probability density. Results are shown for PBE calculations of all the molecules considered [top], and for PBE and PBE0 for select representative molecules [middle and bottom respectively]. Horizontal lines denote average values, except in the case of the rotated molecules, where they correspond to the relaxed molecule. The data include the unmodified extracted molecules, rigidified molecules, and explicitly rotated molecules. \label{fig:valence_quantities}}
\end{figure*}

First, in order to assess the effectiveness of our approach for selecting representative molecules, including the chosen threshold for $J$, PBE calculations were performed for all 500 extracted molecules, as well as the 500 corresponding rigidified molecules. The frontier orbital energies and band gaps are depicted in Fig.~\ref{fig:valence_quantities}a-c alongside the results for the explicitly rotated molecules. Fig.~\ref{fig:valence_quantities}d-f show the same quantities, in the case where only the 42 select molecules are considered (both the unmodified and rigidified versions).  As with the total energy landscape, the representative molecules are able to capture the general trends well, including both the distribution and the average values. In the following we therefore consider the results for the representative molecules only.

\begin{figure}[!ht]
\centering
\begin{subfigure}[t]{0.4\linewidth}
\centering
\includegraphics[width=0.9\textwidth]{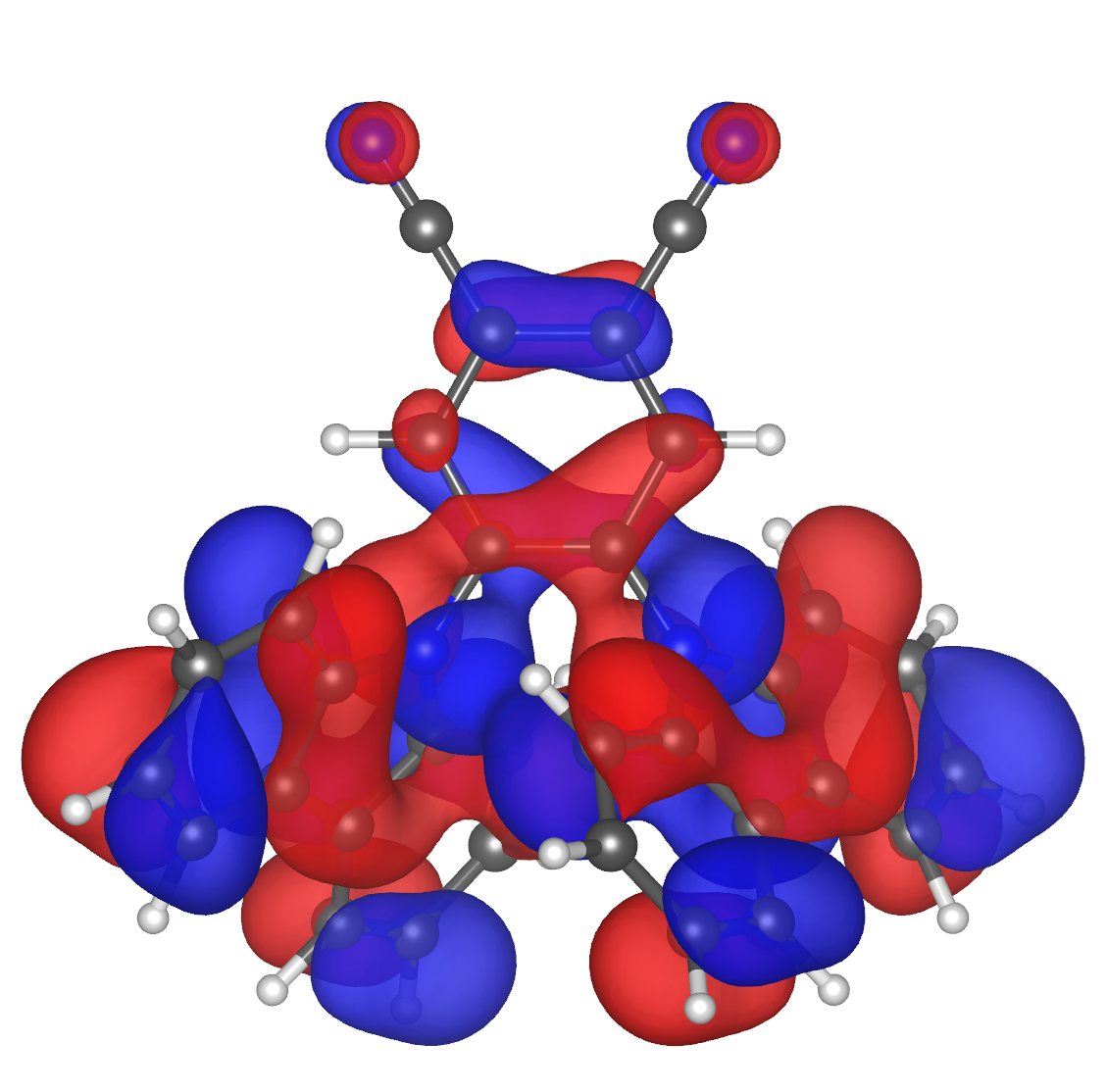}
\caption{HOMO}
\label{fig:homo}
\end{subfigure}
\begin{subfigure}[t]{0.4\linewidth}
\centering
\includegraphics[width=0.9\textwidth]{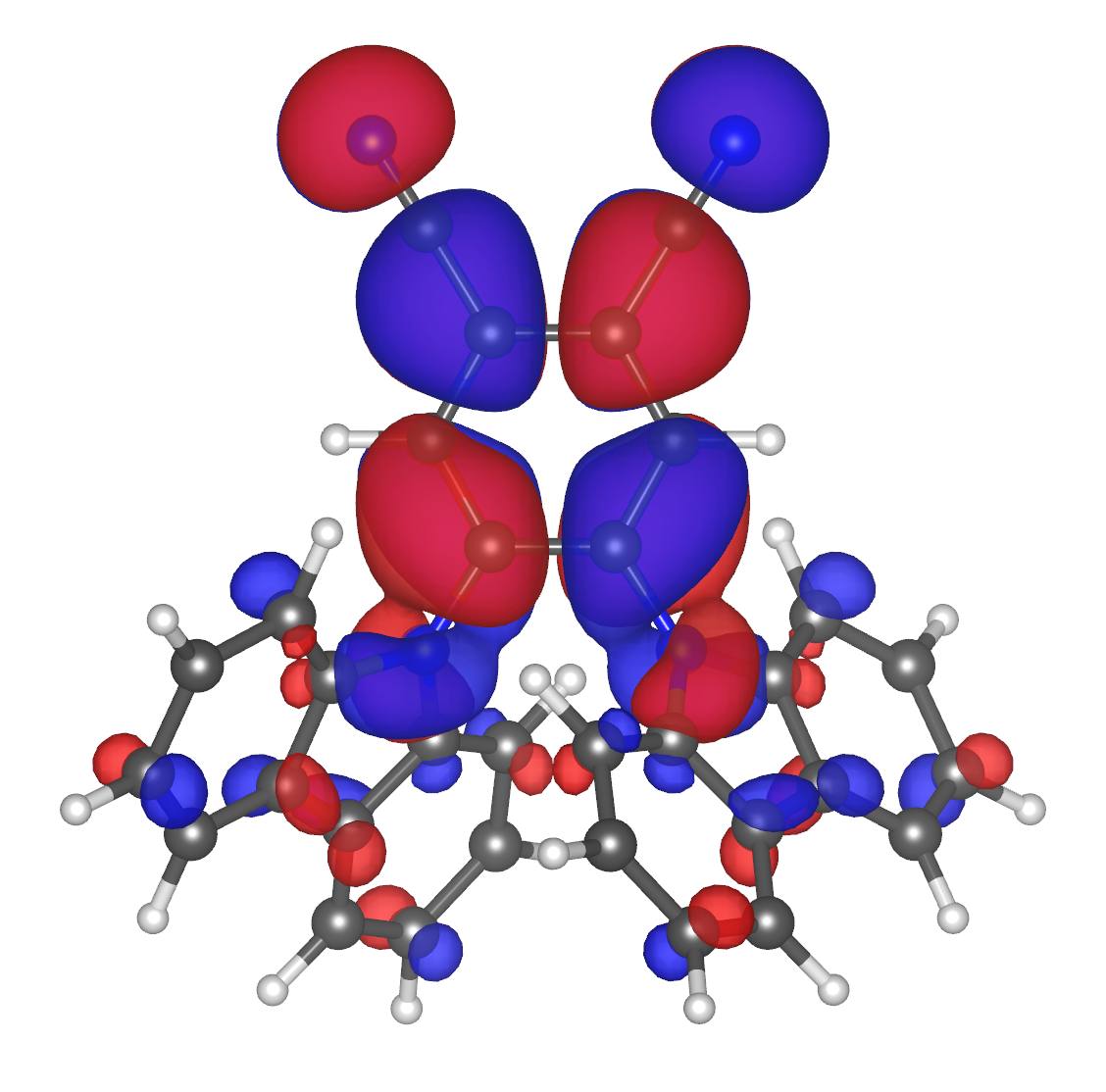}
\caption{LUMO}
\label{fig:lumo}
\end{subfigure}
\caption{Depiction of the PBE-calculated HOMO and LUMO of the relaxed molecule, visualized in VESTA~\cite{Momma2011} using an isosurface value of 0.002~$a_0^{-3/2}$.\label{fig:homo_lumo}}
\end{figure}

Considering first the frontier orbital energies, the highest occupied molecular orbital (HOMO) shows a weaker sensitivity to the torsion angle than the lowest unoccupied molecular orbital (LUMO), so that the trend in band gap is more similar to that of the LUMO. As shown in Fig.~\ref{fig:homo_lumo}, the HOMO is more concentrated on the carbazole units, while the LUMO is centred on the phenylene core, suggesting that this central core is more sensitive to changes in torsion angle, as will be discussed in more detail in the context of core states. The HOMO and LUMO also follow different trends with respect to the change in torsion angle, with the HOMO showing a local maximum at $90^{\circ}$ and two local minima around 50$^{\circ}$ and 130$^{\circ}$.  In contrast, both the LUMO and the band gap have a minimum at $90^{\circ}$.

Considering next both the unmodified and rigidified extracted molecules, the trend with respect to average torsion angle is less evident for all three quantities. In other words, the effects of independently varying torsion angles, and to a lesser extent, variation in bond length between subfragments, lead to a larger spread in values. Indeed, the band gap of the unmodified extracted values varies over a range of 0.8~eV compared to 0.6~eV for the rotated molecules, a larger range than that over which the singlet-triplet splitting was previously found to vary~\cite{Olivier2017}. In all cases the variation is large enough such that the effects of thermal fluctuations cannot be neglected. In addition, there is a net shift in the average value away from the values for the relaxed molecule, which reflects the distribution of angles present in the original structure.  On the other hand, when going from the rigidified to the unmodified extracted molecules, the changes are much smaller, indicating that the effect of non-rigidity is relatively unimportant.

The above results are with PBE, however semi-local functionals such as PBE are expected to significantly underestimate the band gap compared to experiment.  Therefore, the frontier orbital energies of the representative molecules were also calculated using PBE0, the results for which are depicted in Fig.~\ref{fig:valence_quantities}g-i.  Unsurprisingly, there is a net shift in values of around -0.8~eV for the HOMO and 0.8~eV for the LUMO, leading to a shift in the PBE0 gaps of around 1.6~eV relative to the PBE values. However, the overall trends are similar, as is further evidenced in Fig.~S5 in the Supplementary Information.  The external environment can have an important effect on the measured values of the HOMO, LUMO and band gap, contributing to the wide range of experimental values found in the literature. In addition, even in the case where the same solvent (toluene) is used, significant differences exist~\cite{Kim2017,Zhang2019}. Nonetheless, the reported band gaps lie between the PBE and PBE0 values for the relaxed molecule (2.1 and 3.7~eV, respectively).  This is also the case for the measured HOMO and LUMO values.  Although the differences between experimental and theoretical setups limit the extent to which agreement should be expected, it is nonetheless interesting to note that the average PBE0 gap over the extracted molecules (3.5~eV) lies closer to the experimental values than the PBE0 value for the relaxed molecule (3.7~eV), as do the average HOMO and LUMO values.

\subsubsection{Valence Spectra}

Having discussed the frontier orbital energies, we now consider the full range of occupied states. These are also compared to measured valence XPS data, by applying appropriate photoionisation cross-sections to the PDOS, as described in Section~\ref{sec:dft}. In order to verify that the use of representative molecules remains a reasonable approximation, the DOS was calculated using PBE for all molecules. The averaged DOS for the full set of molecules is indistinguishable from the average over the representative molecules, for both the rigidified and unmodified extracted molecules (see Fig.~S6 in the Supplementary Information).  Therefore, as in the previous section we only consider the 42 representative molecules in the following.

Before investigating the effects of disorder, we first consider the unweighted PDOS for the relaxed molecule, which is depicted in Fig.~\ref{fig:dos_disorder}a for PBE0 (see Fig.~S7a in the Supplementary Information for PBE). The PBE0 and PBE PDOS exhibit similar features and relative peak heights, with the key difference being that the PBE0 DOS has a larger overall bandwidth. This also affects the DOS close to the valence band edge, where a shoulder in the PBE DOS becomes a separate peak in the PBE0 DOS.  On the other hand, the projection onto the different atomic orbitals is largely unaffected by the functional. The cross-section weighted PDOS for PBE0 as well as the experimental valence spectrum are shown in Fig.~\ref{fig:dos_disorder}b (see Fig.~S7b in the Supplementary Information for PBE). The experiment shows a number of clearly discernible features and the PBE0 simulated spectra match experiment much more closely than PBE regarding both the total width of the valence states as well as their relative positions.  Indeed, the PBE0 spectrum shows very good agreement with the experimental binding energy positions of the valence features. The relative theoretical peak heights are in less good agreement with experiment, in large part due to an overestimation of the N~$s$ contribution. This is most likely a limitation of the employed projection scheme; in future work it could therefore be interesting to test alternative projection schemes.

\begin{figure}[!ht]
\centering
\includegraphics[scale=0.37]{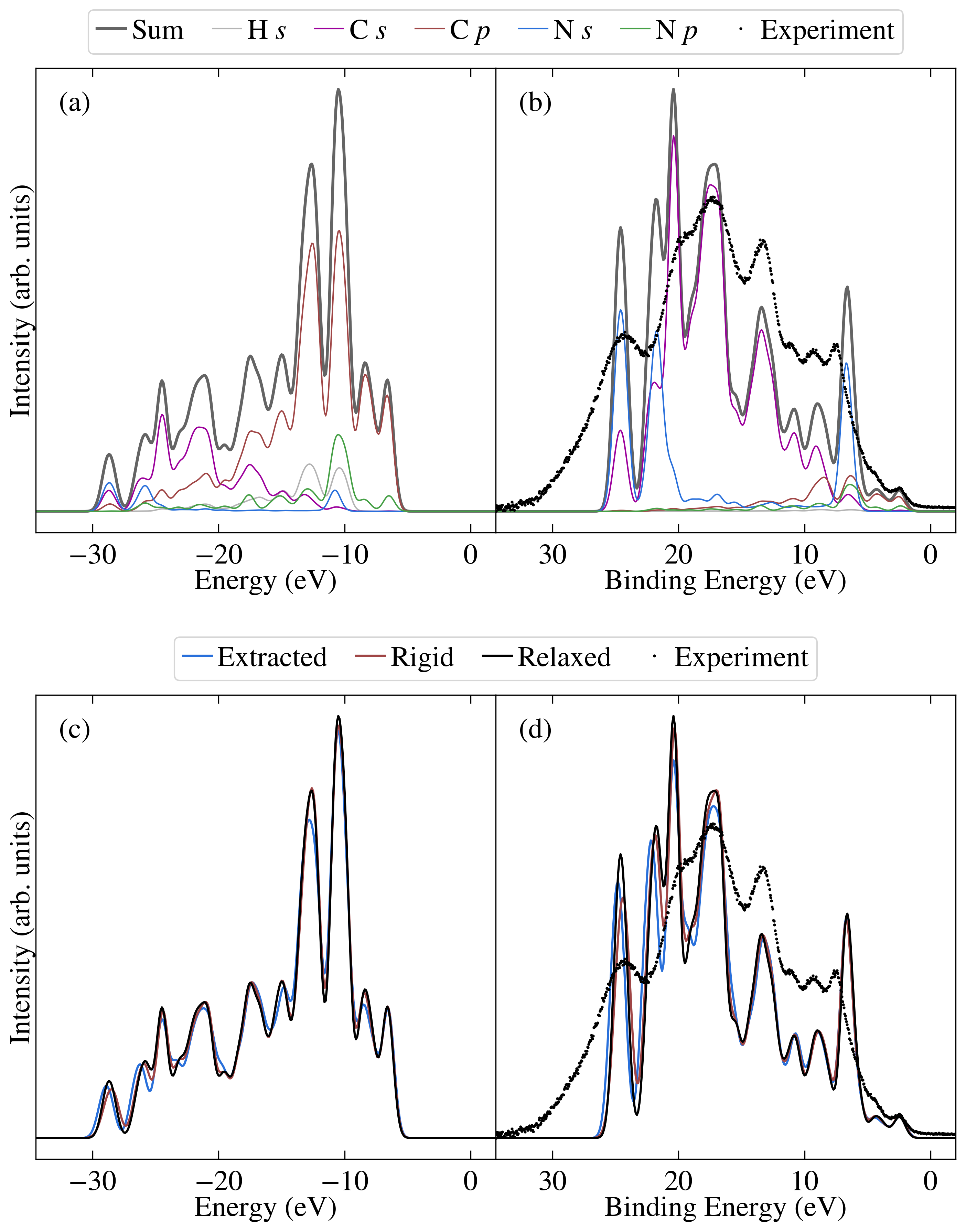}
\caption{PBE0-calculated occupied projected density of states for the relaxed molecule, both (a) unweighted and (b) photoionisation cross-section weighted. Also shown are the sum over contributions for the relaxed molecule, the average over select unmodified extracted molecules, and the average over rigidified molecules, for both (c) unweighted and (d) photoionisation cross-section weighted. The weighted spectra have been aligned to experiment.  \label{fig:dos_disorder}}
\end{figure}

We now consider the effects of disorder, for which the average extracted and rigid results are compared to the relaxed molecule in Fig.~\ref{fig:dos_disorder}c and d for PBE0 (and Fig.~S7c and d in the Supplementary Information for PBE). Although the unweighted PDOS shows some sensitivity to average torsion angle across the full range of energies (see also Fig.~S8 in the Supplementary Information), for the weighted PDOS significant deviations between the averaged disordered spectra are only present for higher binding energies (above 15~eV). These changes are not attributable to one single factor, with the extracted, rigid and relaxed spectra deviating to different extents for different peaks. In other words, as well as showing differences in sensitivity to average torsion angle, some peaks are more sensitive to the effects of non-rigidity than others. This also proves to be the case for the core states, as will be demonstrated in the following.  The spectrum was measured for crystalline 2CzPN, where it is not clear to what extent the range of accessible torsion angles might be limited by intermolecular interactions. Nonetheless, the average spectra over extracted molecules is in slightly better agreement with respect to experiment in terms of peak heights, compared to the relaxed molecule. Although subtle, this suggests that the presence of disorder does indeed have an influence on the spectrum, an assumption which is also examined for the core states.

\subsection{Core States}

\subsubsection{Level of Theory}

The computational cost of calculating core BEs using $\Delta$SCF is significantly higher than for valence quantities, since each C and N atom requires a separate core-excited calculation for each molecule. Furthermore, due to the presence of disorder, the number of atoms to be treated cannot be reduced by relying on symmetry, as could be done for example in an ideal crystal, since each atom may have its own distinctive environment.  In order to keep the computational cost tractable for core BE calculations we therefore do not consider the full set of molecules, only the representative molecules.  Based on the results of Section~\ref{sec:valence}, we expect that this should give sufficient coverage of the potential energy landscape.

Given the high computational cost, it is important to balance cost and accuracy when deciding the level of theory.  To this end, core BEs were calculated using both the Koopmans' and $\Delta$SCF approaches, since unlike $\Delta$SCF, Koopmans' BEs can be calculated using a single (all electron) calculation per molecule. Although there is often a large shift in absolute BEs between the two approaches, in this work we focus on relative BEs, and thus this shift can be ignored. More importantly, however, final state effects, which are not captured by Koopmans', can sometimes lead to a qualitatively wrong BE order, as shown for example in the aromatic amino acids~\cite{Regoutz2020}. It is therefore important to determine how significant such effects are for 2CzPN.  Aside from the expected large systematic offset, the Koopmans' and $\Delta$SCF BEs show similar trends for most atoms, although the Koopmans' BEs are more spread out compared to $\Delta$SCF (see Fig.~S9 in the Supplementary Information). However, the C atoms in the cyano group follow a notably different trend, with their BEs being significantly underestimated using the Koopmans' approach relative to the BEs associated with other C atoms. Therefore, all results in the following employ $\Delta$SCF.

Although prohibitively expensive to calculate $\Delta$SCF BEs for all molecules using PBE0, a comparison was nonetheless performed for the relaxed molecule, to determine to what extent the results are sensitive to the functional.  In addition to a systematic offset, the C~$1s$ BEs are more spread out for PBE0 than PBE, however, the PBE and PBE0 results are well correlated, with the cyano C atoms being much less sensitive to the choice of functional than the inclusion of final state effects (see Fig.~S10 in the Supplementary Information).  In other words, PBE represents a reasonable approximation for C~$1s$, and so except where otherwise stated, all results in the following are for PBE. For N~$1s$ the BEs are more strongly affected by the choice of functional than C~$1s$, with the separation between the two pairs of BEs being 1.7~eV for PBE compared to 1.4~eV for PBE0.  In addition to the relaxed molecules, N~$1s$ BEs were therefore also calculated for three additional rotated molecules, as shown in Table~\ref{tab:n_bes}. Importantly, the difference between the PBE and PBE0 separations is consistently 0.3~eV, so that although the relative BEs are sensitive to the choice of functional, the trends for PBE and PBE0 are nonetheless expected to be similar.

\begin{table}[h]
\small
\caption{N~$1s$ binding energies and binding energy differences, $\Delta$, in eV, calculated for select explicitly rotated molecules using PBE and PBE0, and compared to experiment. The experimental values have an estimated error of $\pm$0.1~eV.\label{tab:n_bes}}
\begin{tabular*}{0.48\textwidth}{@{\extracolsep{\fill}}l rrr rrr }
\hline
 & \multicolumn{3}{c}{PBE} &   \multicolumn{3}{c}{PBE0} \\
\cline{2-4}\cline{5-7}\\[-1.5ex]
& N$_1$ & N$_2$ & $\Delta$ & N$_1$ & N$_2$ & $\Delta$ \\
\cline{1-1}\cline{2-2}\cline{3-3}\cline{4-4}\cline{5-5}\cline{6-6}\cline{7-7}\\[-1.5ex]
$40.0^\circ$ & 413.0 & 411.2 & 1.8 & 418.6 & 417.1 & 1.5\\
$51.9^\circ$ (relaxed) & 413.0 & 411.3 & 1.7 & 418.5 & 417.2 & 1.4\\
$65.0^\circ$ & 412.9 & 411.4 & 1.5 & 418.5 & 417.3 & 1.1\\
$90.0^\circ$ & 412.8 & 411.8 & 1.0 & 418.4 & 417.7 & 0.7 \\
\cline{1-1}\cline{2-2}\cline{3-3}\cline{4-4}\cline{5-5}\cline{6-6}\cline{7-7}\\[-1.5ex]
Experiment & 400.5 & 399.5 & 1.0 \\
\hline
\end{tabular*}
\end{table}

\subsubsection{Experimental Spectra}

\begin{figure}[!ht]
\centering
\includegraphics[scale=0.37]{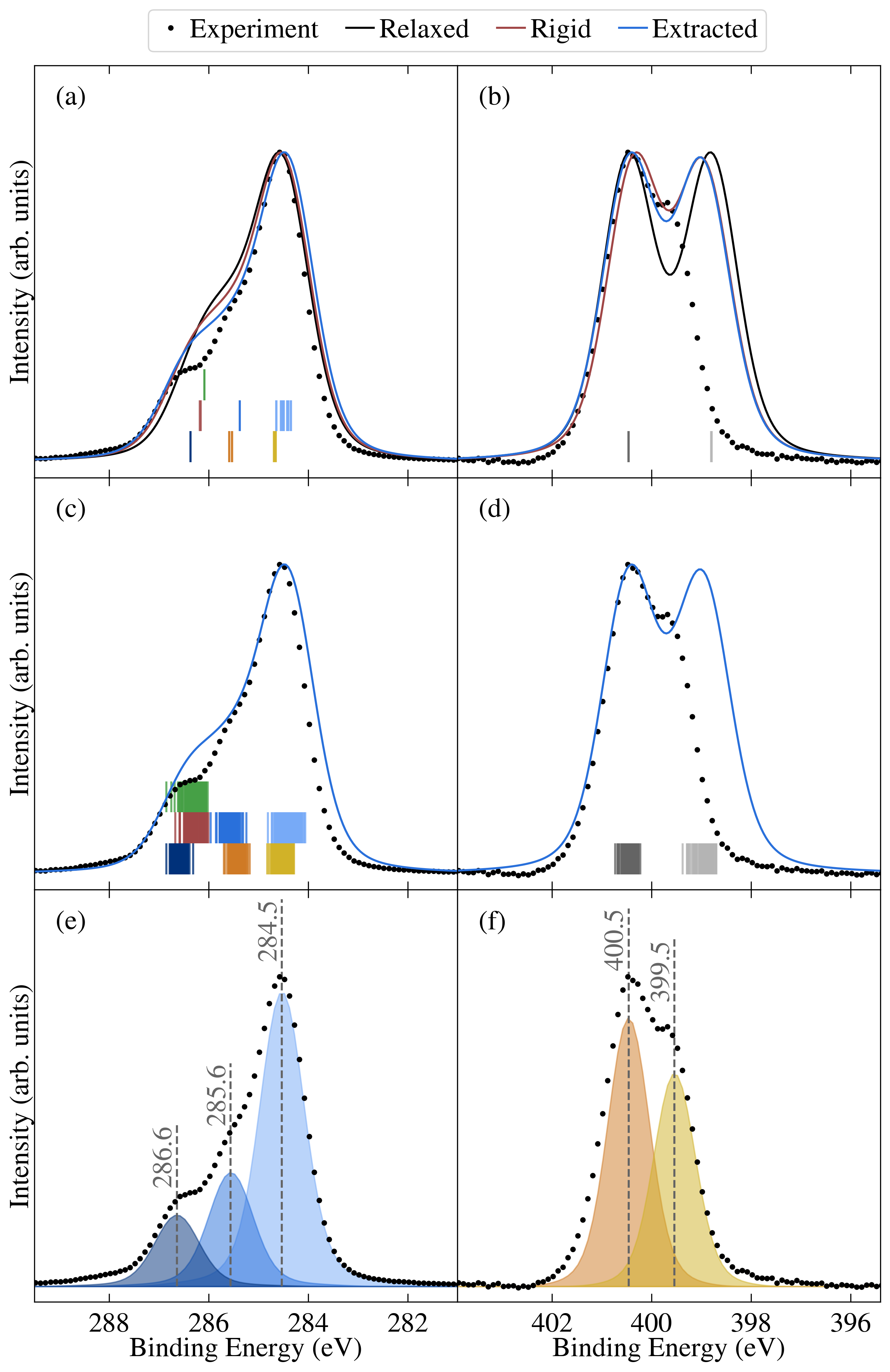}
\caption{Comparison between experimental (dots) and theoretical (lines) core spectra, including the relaxed molecule, the average over select unmodified extracted molecules, and the average over rigidified molecules, calculated using PBE for (a) and (c) C~$1s$ and (b) and (d) N~$1s$.  The colours correspond to different atomic sites, as depicted in Fig.~\ref{fig:core_colours}. Also shown are vertical lines denoting the calculated binding energies for the relaxed molecule in (a) and (b) and the unmodified extracted molecules in (c) and (d). Theoretical results have been aligned to experiment as described in section~\ref{sec:dft}. Also shown are the experimental peak fits for (e) C~$1s$ and (f) N~$1s$. \label{fig:core}}
\end{figure}

X-ray photoelectron spectra of the C~$1s$ and N~$1s$ core levels of 2CzPN are presented in Fig.~\ref{fig:core}. Figs.~\ref{fig:core}e and f show the main chemical state contributions identifiable from peak fit analysis. The C~$1s$ spectrum exhibits three features at 284.5, 285.6 and 286.6~eV. Based on the theoretical BE values, these three peaks can be assigned to specific C atoms within the 2CzPN molecule as follows. The lowest BE contribution corresponds to the C$_1$ and C$_2$ atoms of the two carbazole units that are only bound to other C and H atoms, commensurate with basic arguments of chemical shift in carbon spectra. In contrast, the peak at 285.6~eV, assigned to C$_3$ and C$_4$ atoms, and the peak at 286.6~eV, associated with the C$_5$, C$_6$ and C$_7$ atoms, show BE positions not fully expected from basic arguments and experimental experience. The BE position of the C$_3$ and C$_5$ atoms are considerably higher than what is expected from atoms bound only to other C and H atoms. This observation is commensurate with our previous work on aromatic amino acids, where next-nearest neighbour effects as well as the influence of aromatic rings on BEs are in effect~\cite{Regoutz2020}. The relative peak intensities in the experiment match well with the expected atomic ratios for the C$_1$ and C$_2$ peak (61~rel.\ at.\% from experiment and 62.5\% from molecular structure) as well as with the sum of the two higher BE contributions (39~rel.\ at.\% from experiment and 37.5\% from molecular structure). It is worth noting, that whilst the latter two features would be expected to appear in a 1:1 intensity ratio from the theoretical calculations, the experiment shows relative contributions of 24 and 15~rel.\ at.\%. 

The N~1$s$ core level spectrum has two major contributions at 399.5 and 400.5~eV assigned to the N$_1$ and N$_2$ atoms of the molecular structure, respectively. The relative contributions of the two N environments from peak fit analysis is 44 and 56~rel.\ at\%, respectively. In addition to the discussed expected chemical environments of both C and N, the core level spectra show small additional contributions from surface species , expected for samples handled in air.

\subsubsection{Influence of Disorder}

\begin{figure*}[!ht]
\centering
\includegraphics[scale=0.37]{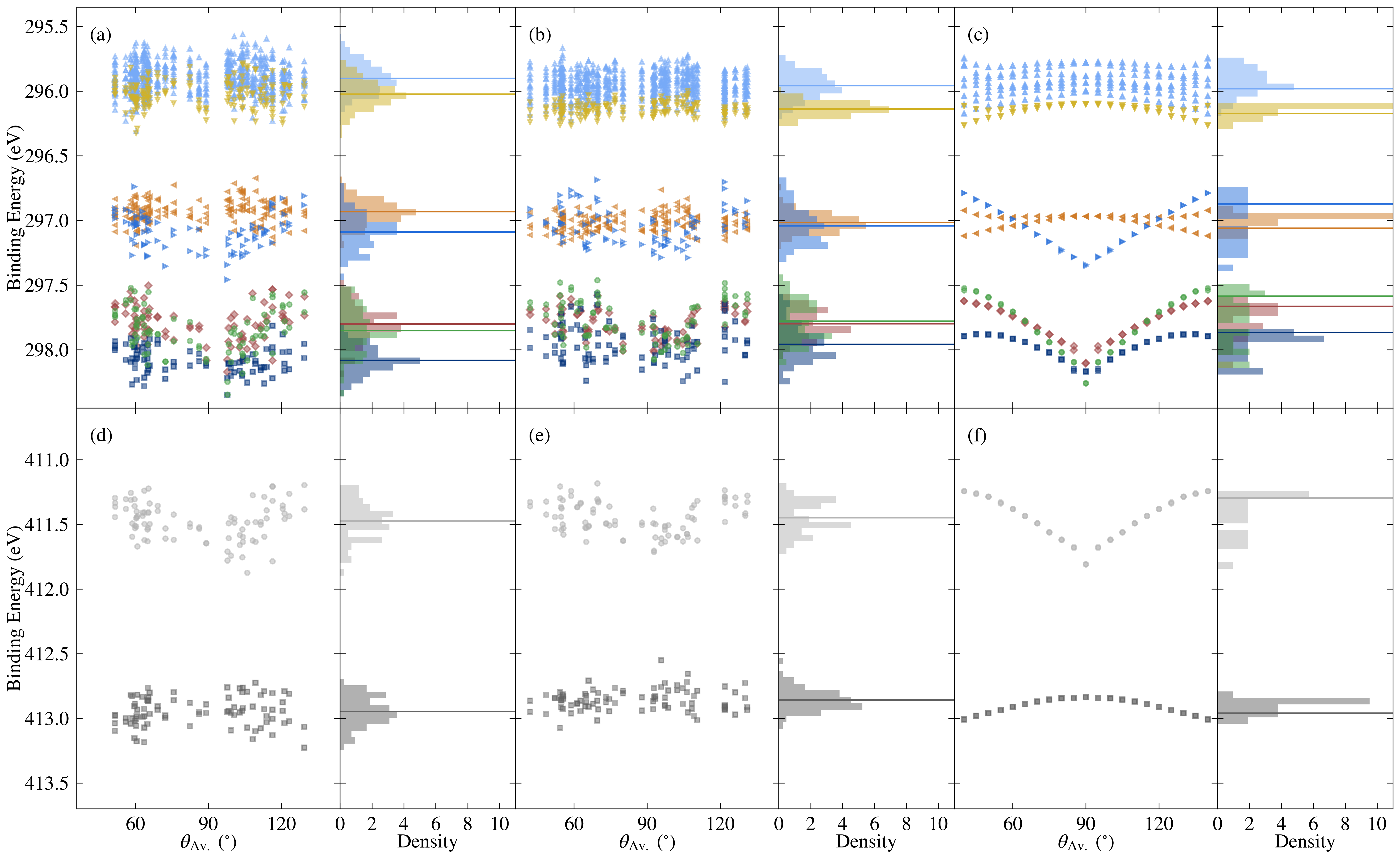}
\caption{Variation in core binding energies with respect to the average torsion angle $\theta_{\mathrm{Av.}}$, as well as the probability density, for C~$1s$ [top] and N~$1s$ [bottom], where the colours correspond to different atomic sites, as depicted in Fig.~\ref{fig:core_colours}. Horizontal lines denote average values, except in the case of the rotated molecules, where they correspond to the relaxed molecule. The data include (a) and (d) the unmodified extracted molecules, (b) and (e) the rigidified molecules, and (c) and (f) the explicitly rotated molecules. \label{fig:core_statistics}}
\end{figure*}

Considering now the effect of the torsion angle on the core state BEs, Fig.~\ref{fig:core_statistics}c and f show the dependence on average torsion angle for the rotated molecules. Although there are variations between both the degree of sensitivity to the torsion angle and the exact shape of the trend, there are two distinct scenarios: the atoms associated with the phenylene core (C$_3$, C$_{5-7}$ and N$_1$) show a maximum at $90^\circ$ and a higher sensitivity to the torsion angle, while the atoms associated with the carbazole units (C$_{1-2}$, C$_4$, N$_2$) show a much shallower minimum at $90^\circ$. This confirms earlier observations that the phenylene core is more sensitive to changes in torsion angle (see Section~\ref{sec:frontier_orbitals}).  The strong sensitivity of the N$_1$ BE to torsion angle leads to a large variation in BE separation between N$_1$ and N$_2$ -- ranging from 1.1 to 1.8~eV (Table~\ref{tab:n_bes}), a variation which is clearly visible in the calculated spectra (see Fig.~S11 in the Supplementary Information).  In contrast, the C~$1s$ theoretical spectra show much more subtle spectral changes with varying torsion angle.

Going beyond the explicitly rotated molecules, Fig.~\ref{fig:core_statistics} also shows the variation in core BEs for the extracted and rigidified molecules.  As with the frontier orbitals, the additional disorder dilutes the strong trend with respect to torsion angle in both cases, leading to wider variations in BEs. The extent to which particular BEs are sensitive to the different levels of disorder varies. For example the N$_1$ BEs show similar spreads for the rotated, rigidified and extracted molecules (0.6, 0.5 and 0.7~eV respectively),  whereas N$_2$ BEs have a spread of 0.2~eV for the rotated molecules, increasing to 0.5~eV for both the rigidified and extracted molecules (see also Fig.~S12 and Table~S1 in the Supplementary Information). The C~$1s$ BEs also show different patterns of spreads, however, there is no obvious trend concerning which atoms are more sensitive to rigidity than others. Despite this variation in behaviour, both the ranges of values and shifts in average relative to the relaxed molecule are the same order of magnitude as for the frontier orbitals.

As shown by comparing Fig.~\ref{fig:core}a and c, the large spreads in C~$1s$ BEs of the extracted molecules (up to 0.8~eV) nonetheless have only a relatively subtle impact on the overall spectrum.  Indeed, while the C~$1s$ spectrum for the relaxed molecule already agrees well with experiment, the higher BE shoulder reduces slightly in intensity going from the relaxed to rigid to extracted molecules, leading to a slightly improved experimental agreement. This is in part due to the spreading out and resulting decreased separation between C$_3$ and C$_5/6$ BEs, which also helps explain the discrepancy between the experimental and theoretical relative contributions of the two higher BE peaks.

On the other hand, the variation in N~$1s$ BEs (up to 0.7~eV) lead to larger variations in the theoretical spectra.  Indeed, as shown in Fig.~\ref{fig:core}b and d, the extracted and rigid molecules give rise to a better match to experiment than the relaxed molecule, although the BE separation is still considerably overestimated.  However, as previously discussed this separation is sensitive to both the functional and the torsion angle.  Indeed, as shown in Table~\ref{tab:n_bes}, the PBE0 separation for the relaxed molecule is 0.4~eV higher than experiment, whereas the PBE value is 0.7~eV higher (see also Fig.~S13 in Supplementary Information). At the same time, the difference between average N$_1$ and N$_2$ PBE-calculated BEs for the extracted molecules reduces to 1.4~eV, compared to 1.7~eV in the relaxed molecule.

The high computational cost prohibits calculating the BEs for all of the extracted molecules  with PBE0.  However, we may instead employ a simplified variation of the so-called spectral warping approach, also known as colour morphing, which has been employed in the context of optical excitations calculated using TDDFT~\cite{Ge2015,Zuehlsdorff2017,Prentice2021}. In this approach, a correction is imposed on semi-local calculations by comparing with representative hybrid functional calculations. In this case, we may simply impose a shift of 0.3~eV on the N$_1$ BEs. The smaller separation of the two peaks in the warped spectrum does indeed lead to a closer match between the theory and experiment, albeit with the overall bandwidth still being slightly overestimated in the simulated spectrum (see Fig.~S14 in the Supplementary Information). As a result, the agreement remains slightly worse than for C~$1s$. However, such a strong sensitivity of the N~$1s$ BEs suggests that they may also be affected by other factors, such as intermolecular interactions with neighbouring molecules. Such effects may also influence the calculated valence spectra, therefore, it would be interesting in future work to assess the impact of going beyond gas phase by including neighbouring molecules. This could be achieved either by extracting a small cluster from the MD trajectory, or by simulating 2CzPN in its crystalline form.

\section{Conclusion}

In this work, theory and experiment are used to investigate the influence of disorder on the core and valence states of the prototypical TADF emitter 2CzPN, by simulating gas phase molecules with varying levels of disorder.  In order to make such calculations tractable, we employ a novel approach for identifying representative molecules drawn from an MD snapshot, which is validated by comparing valence quantities calculated for both the representative molecules and the full data set. The results show that the frontier orbitals exhibit differing trends with respect to torsion angle, with the LUMO also being more sensitive than the HOMO. However, the influence on the simulated X-ray photoelectron spectra is subtle, with the theoretical spectrum already showing good agreement with experiment without taking into account the disorder.

While excited state properties are more directly applicable for understanding the performance of TADF-based OLEDs in devices, core XPS has the advantage that the influence of disorder can be mapped at a more local level.  However, the interpretation of such spectra for large molecules based purely on chemical intuition or reference spectra is challenging.  Indeed, in the case of aromatic molecules such an interpretation may even be qualitatively wrong, due to the importance of both non-nearest neighbour and final state effects.  As such, the theoretical prediction of core binding energies can be invaluable both for interpreting experiments, and for assessing the influence of disorder.  Nonetheless, core BE calculations based on for example DFT are relatively uncommon, and have typically been restricted to small molecules.  In this work, we employ the multi-wavelet-based MADNESS code to calculate core BEs of 2CzPN, in which the $\Delta$SCF approach has recently been implemented, and which is both precise and computationally efficient enough to simulate large molecules.  The results show that both C~$1s$ and N~$1s$ BEs are sensitive to disorder, with some sites more strongly influenced by factors such as varying torsion angle than others. The theoretical C~$1s$ spectrum shows excellent agreement with experiment, and, like the valence XPS, is relatively insensitive to disorder. The experimental spectrum is less well predicted for N~$1s$, while it is also found to be more sensitive to both the employed exchange correlation functional and the disorder. Our approach could be extended in future to also consider environmental effects, by going beyond gas phase to the treatment of e.g.\ clusters of molecules. 

This work demonstrates the importance of taking into account disorder in TADF materials, and the resulting implications for providing an accurate picture of their electronic structure.  A similar approach could be used in future to investigate other TADF emitters.  Furthermore, beyond the implications for TADF-based OLEDs, this work also demonstrates the benefits of using theory to aid in the interpretation of experimental spectra, especially for large aromatic molecules where a reliable interpretation cannot be achieved purely by relying on comparison to reference spectra.

\section*{Conflicts of interest}
There are no conflicts to declare.

\section*{Acknowledgements}
LER and MS acknowledge support from an EPSRC Early Career Research Fellowship (EP/P033253/1).
NKF acknowledges support from the Engineering and Physical Sciences Research Council (EP/L015277/1).
AR acknowledges support from the Analytical Chemistry Trust Fund for her CAMS-UK Fellowship. 
Calculations were performed on the Imperial College High Performance Computing Service and the ARCHER2 UK National Supercomputing Service (https://www.archer2.ac.uk). 
LER and MS would like to thank Yoann Olivier for providing the MD snapshot and for useful discussions. 

Data associated with this paper, including a Jupyter notebook demonstrating the workflow for identifying representative molecules, are available at \url{https://gitlab.com/lratcliff/2czpn}.


\balance

\bibliography{refs} 
\bibliographystyle{rsc} 

\end{document}


\title{{\LARGE Supplementary Information}\\
Probing Disorder in 2CzPN using Core and Valence States}
\date{\today}

\author{Nathalie K. Fernando}
\affiliation{Department of Chemistry, University College London, 20 Gordon Street, London WC1H 0AJ, United Kingdom}

\author{Martina Stella}
\affiliation{Department of Materials, Imperial College London, London SW7 2AZ, United Kingdom}
\affiliation{The Abdus Salam International Centre for Theoretical Physics, Condensed Matter and Statistical Physics, 34151 Trieste, Italy}

\author{William Dawson}
\affiliation{RIKEN Center for Computational Science, Kobe, Japan}

\author{Takahito Nakajima}
\affiliation{RIKEN Center for Computational Science, Kobe, Japan}

\author{Luigi Genovese}
\affiliation{Univ.\ Grenoble Alpes, CEA, IRIG-MEM-L\_Sim, 38000 Grenoble, France}

\author{Anna Regoutz}
\affiliation{Department of Chemistry, University College London, 20 Gordon Street, London WC1H 0AJ, United Kingdom}

\author{Laura E. Ratcliff}
\email{laura.ratcliff@bristol.ac.uk}
\affiliation{Department of Materials, Imperial College London, London SW7 2AZ, United Kingdom}
\affiliation{Centre for Computational Chemistry,
School of Chemistry, University of Bristol, Bristol BS8 1TS,
United Kingdom}

\maketitle

\section{Atomic Structure}

\begin{figure}[h]
\centering
\includegraphics[scale=0.37]{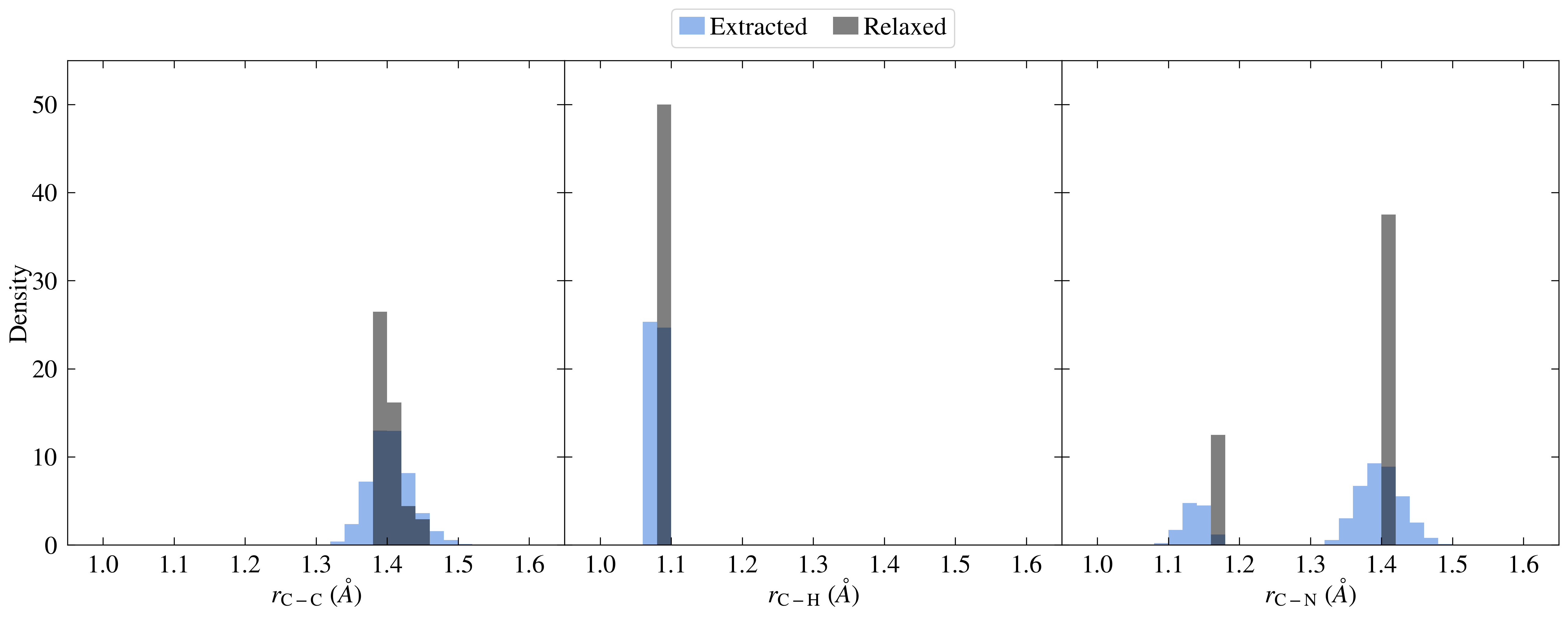}
\caption{Distribution of bond lengths in the unmodified extracted molecules, compared to the relaxed molecule. \label{fig:bonds}}
\end{figure}

\begin{figure}[h]
\centering
\includegraphics[scale=0.37]{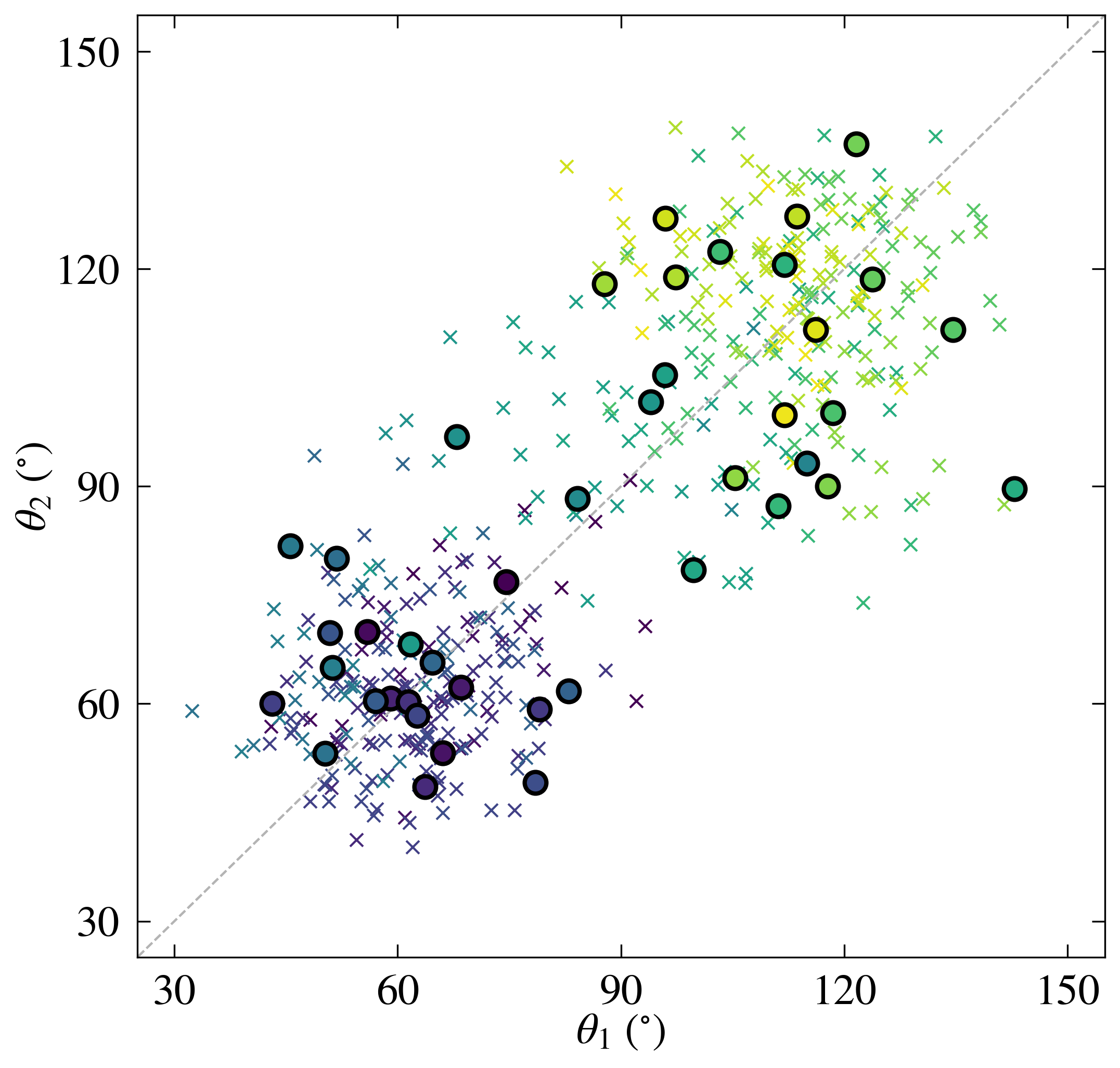}
\caption{Depiction of the torsion angles for each molecule, when grouped by similar molecules (`x'), where the colours correspond to the particular cluster, as depicted in the main paper. Also shown are the select representative molecules from each cluster (`o'). \label{fig:selected_molecules}}
\end{figure}

\clearpage

\section{Valence States}

\begin{figure*}[ht]
\centering
\includegraphics[scale=0.37]{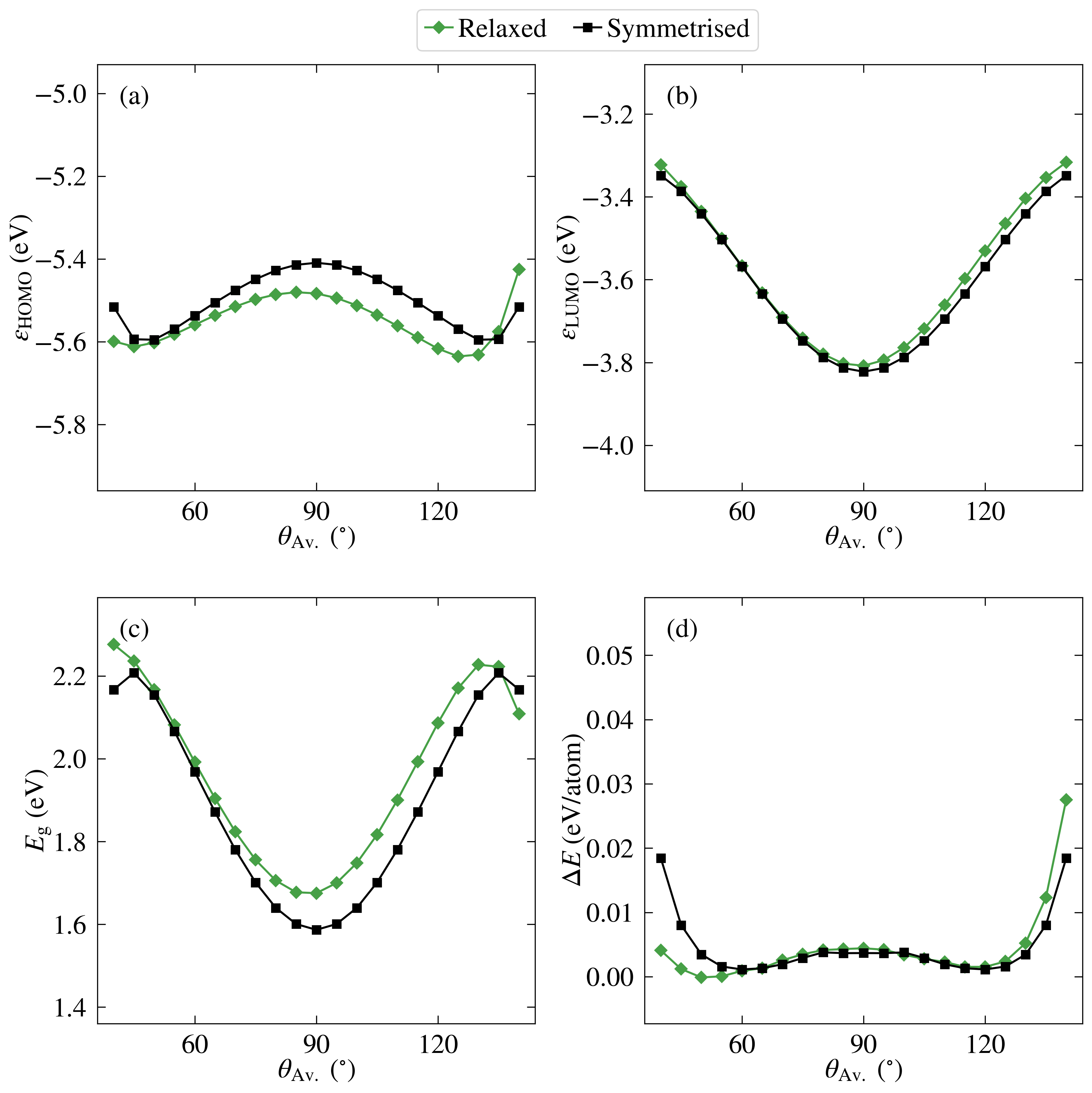}
\caption{Comparison between various quantities calculated for rotated molecules generated by rotating the relaxed molecule without imposing any constraints (`Relaxed'), and by rotating a molecule which has been relaxed with imposed symmetry `Symmetrised'.\label{fig:sym_rot}}
\end{figure*}

\begin{figure*}[ht]
\centering
\includegraphics[scale=0.37]{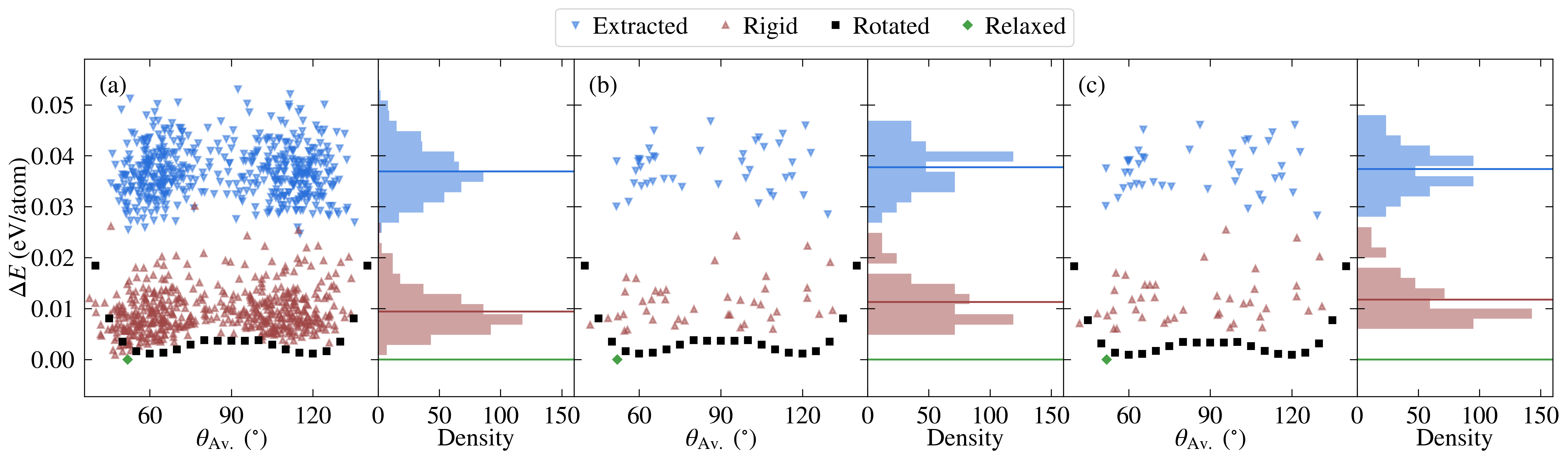}
\caption{Variation in total energy relative to the relaxed structure, $\Delta E$, with the average torsion angle, $\theta_{\mathrm{Av.}}$, as well as the probability density for (a) PBE calculations of all the molecules considered, and for (b) PBE and (c) PBE0 for select representative molecules. Horizontal lines denote average values, except in the case of the rotated molecules, where they correspond to the relaxed molecule. The data include the unmodified extracted, the rigidified molecules, and the explicitly rotated molecules. \label{fig:total_energies}}
\end{figure*}

\begin{figure}[h]
\centering
\includegraphics[scale=0.37]{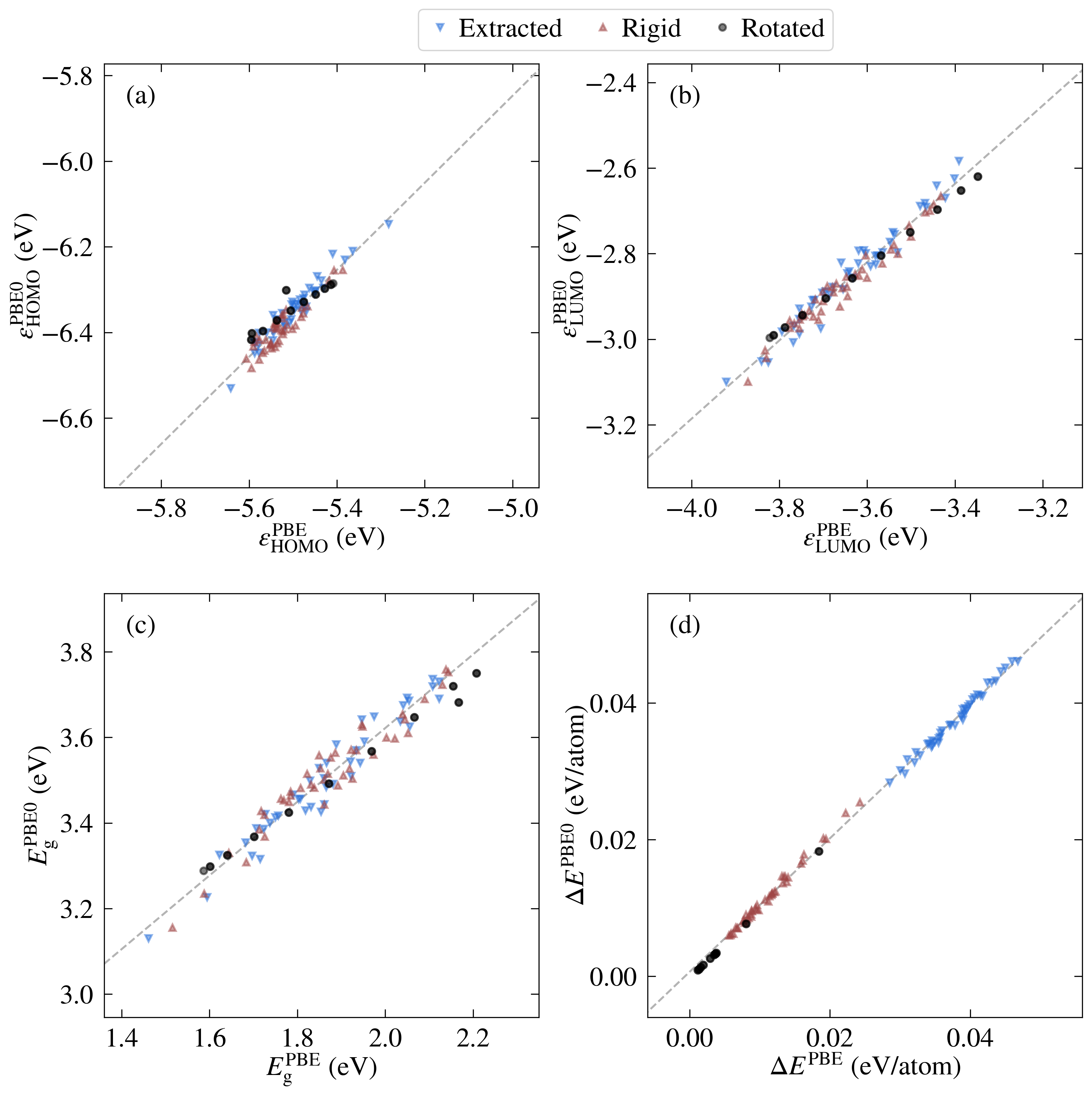}
\caption{Comparison between PBE and PBE0-calculated values for (a) the highest occupied molecular orbital energy, $\varepsilon_{\mathrm{HOMO}}$, (b) the lowest unoccupied molecular orbital energy, $\varepsilon_{\mathrm{LUMO}}$, (c) the band gap, $E_{\mathrm{g}}$, and (d) total energies relative to the relaxed molecule, $\Delta E$.  The data include the unmodified extracted, the rigidified molecules, and the explicitly rotated molecules. \label{fig:valence_pbe0}}
\end{figure}

\clearpage

\begin{figure}[h!]
\centering
\includegraphics[scale=0.37]{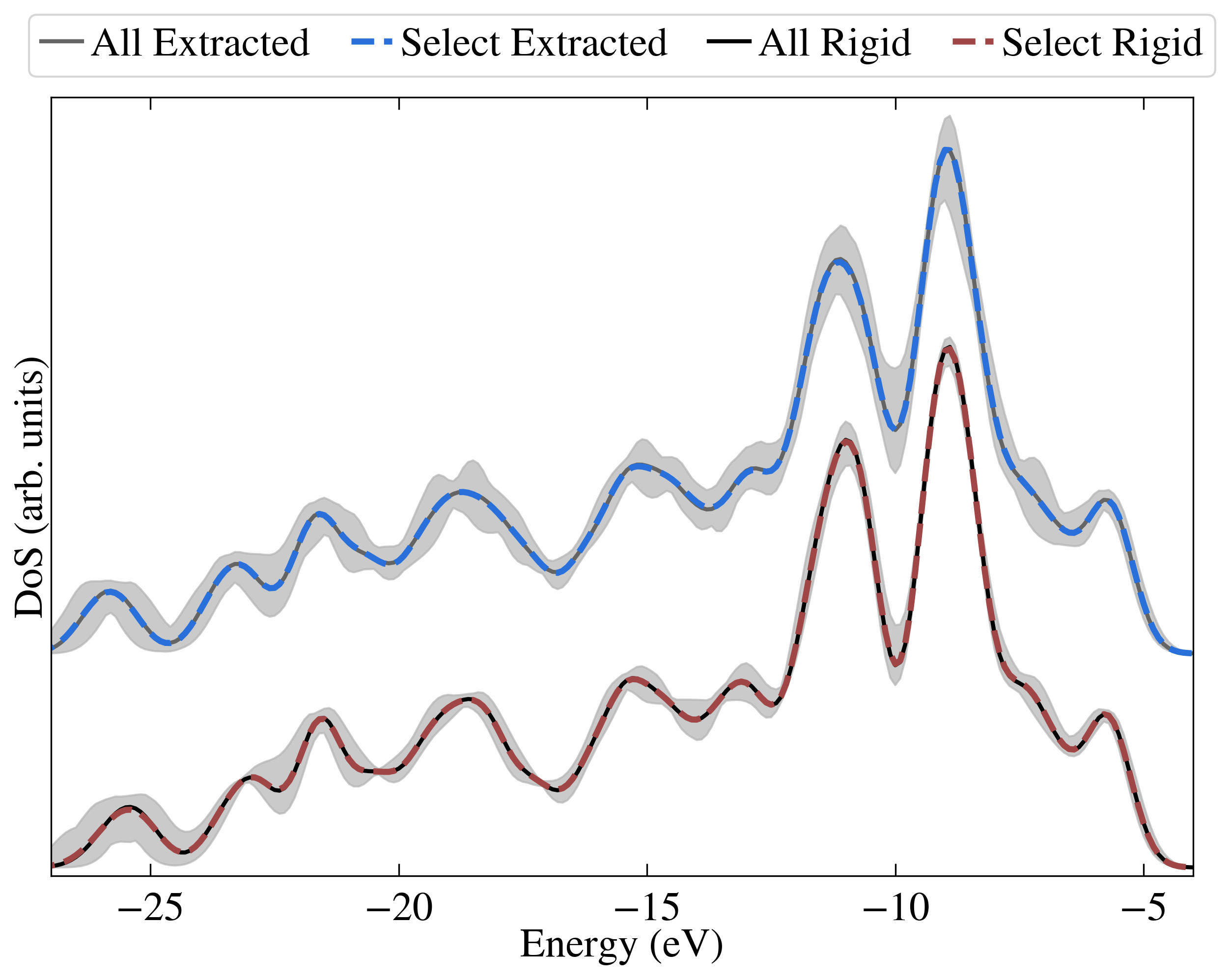}
\caption{Total density of states calculated as an average over all the unmodified extracted (rigidified) molecules, as well as an average over select representative unmodified extracted (rigidified) molecules. The light grey area indicates the variation in calculated densities of states across all unmodified extracted (rigidified) molecules. Calculations employed PBE. \label{fig:select_dos}}
\end{figure}

\begin{figure}[h!]
\centering
\includegraphics[scale=0.37]{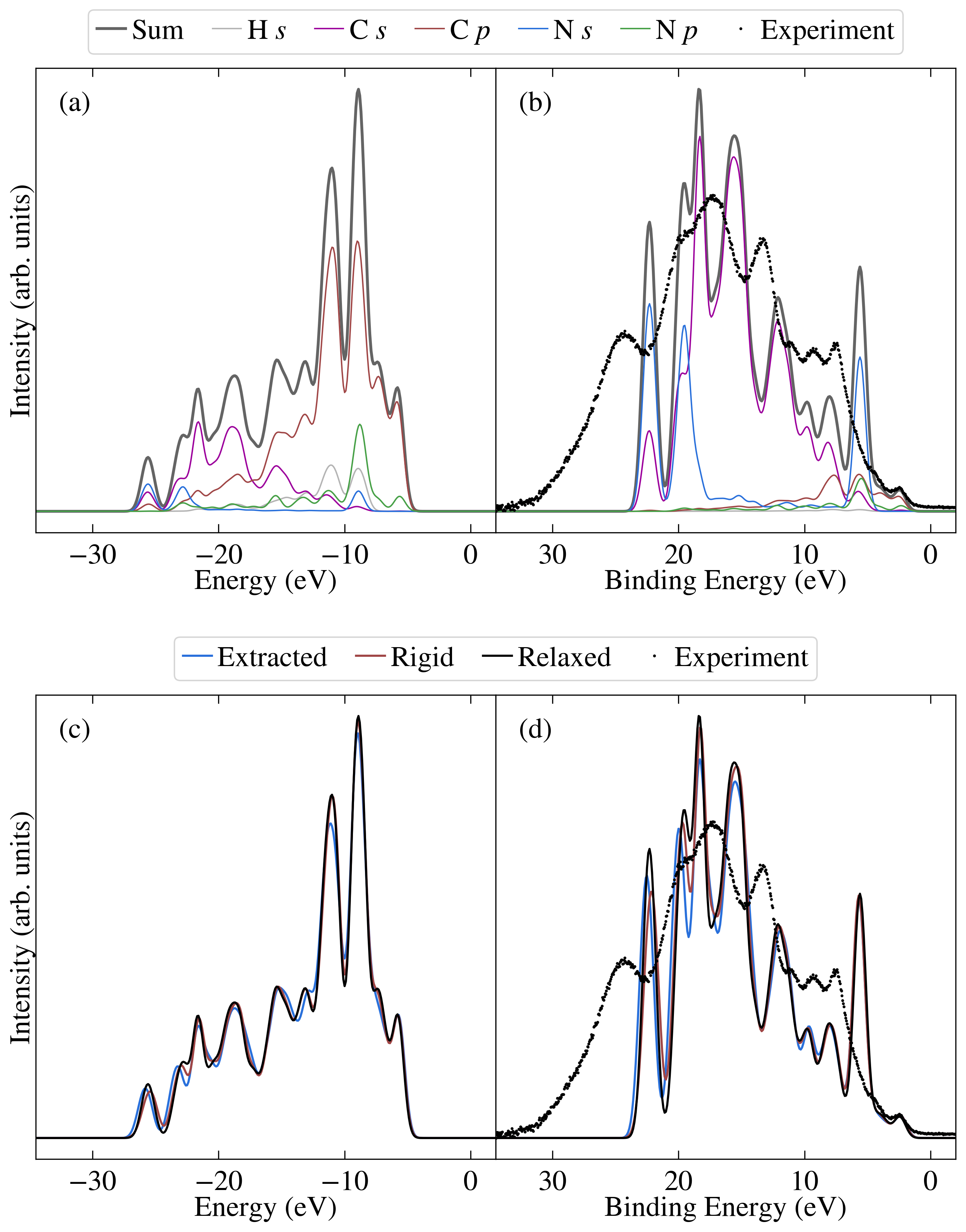}
\caption{PBE-calculated occupied projected density of states for the relaxed molecule, both (a) unweighted and (b) cross-section weighted. Also shown are the sum over contributions for the relaxed molecule, the average over select unmodified extracted molecules, and the average over rigidified molecules, for both (c) unweighted and (d) cross-section weighted. The weighted spectra have been aligned to experiment.  \label{fig:dos_disorder}}
\end{figure}

\begin{figure}[h]
\centering
\includegraphics[scale=0.37]{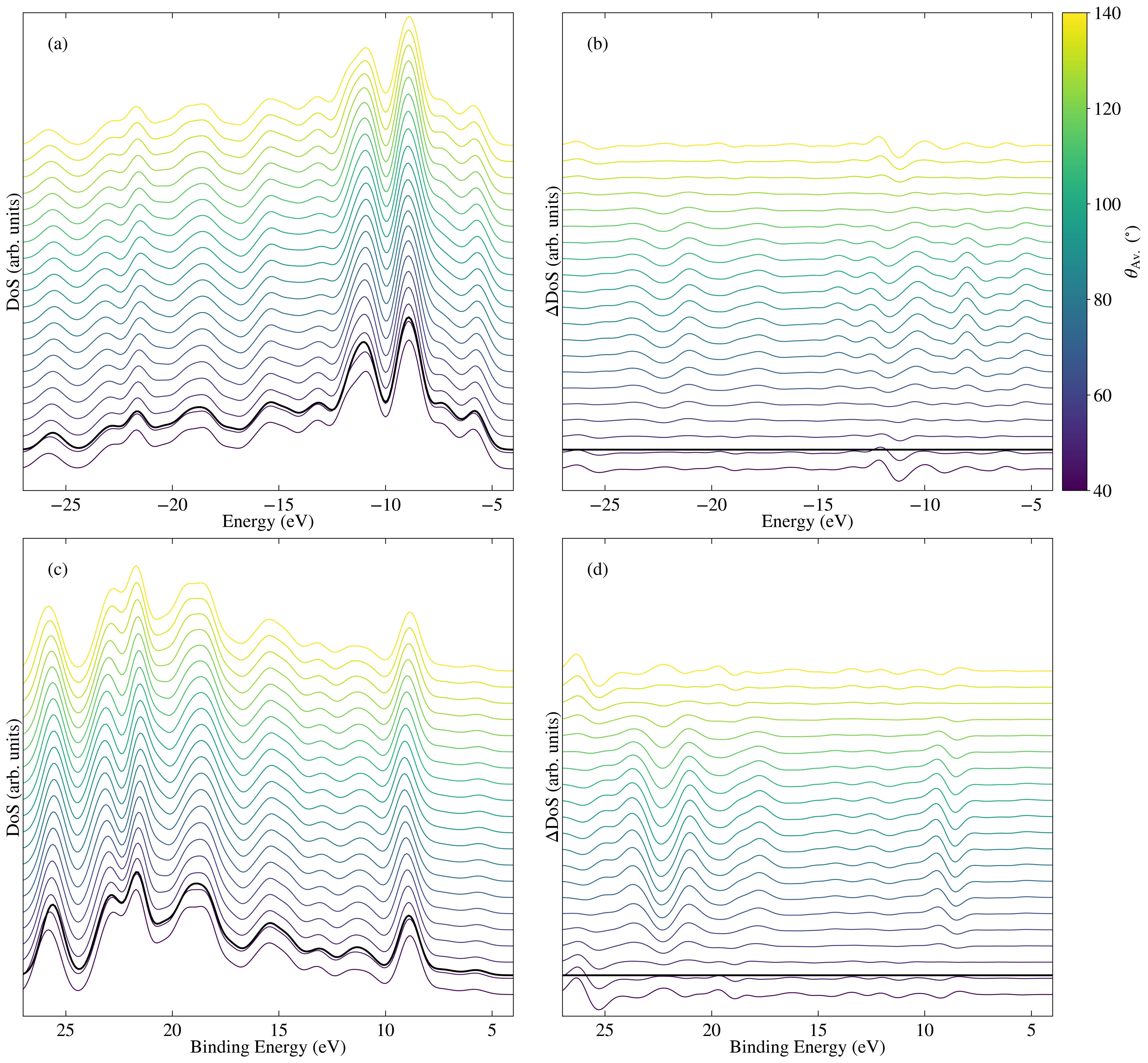}
\caption{Variation in the PBE-calculated total density of states with average torsion angle, for the explicitly rotated molecules, where the relaxed molecule is highlighted with thick black lines. Plots are shown for both (a) the unweighted and (c) the weighted density of states. Also shown is the difference in density of states with respect to the relaxed molecule for (c) the unweighted and (d) the weighted density of states. \label{fig:theta_dos}}
\end{figure}

\clearpage

\section{Core States}

\begin{figure}[h]
\centering
\includegraphics[scale=0.37]{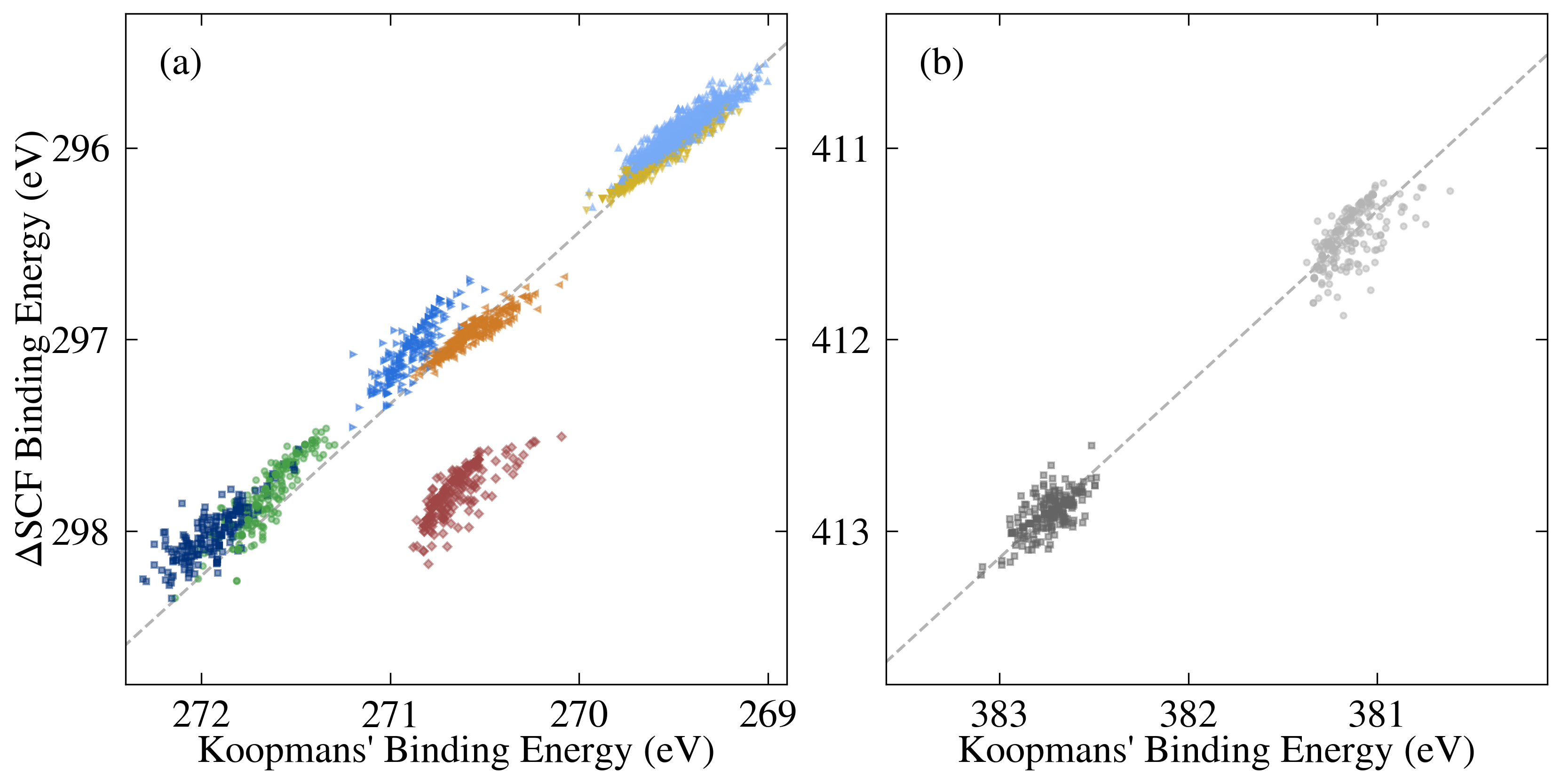}
\caption{Comparison between binding energies calculated using Koopmans' and $\Delta$SCF for (a) C~$1s$ and (b) N~$1s$, where the colours refer to the different atomic sites (see main paper).\label{fig:koopmans}}
\end{figure}

\begin{figure}[h]
\centering
\includegraphics[scale=0.37]{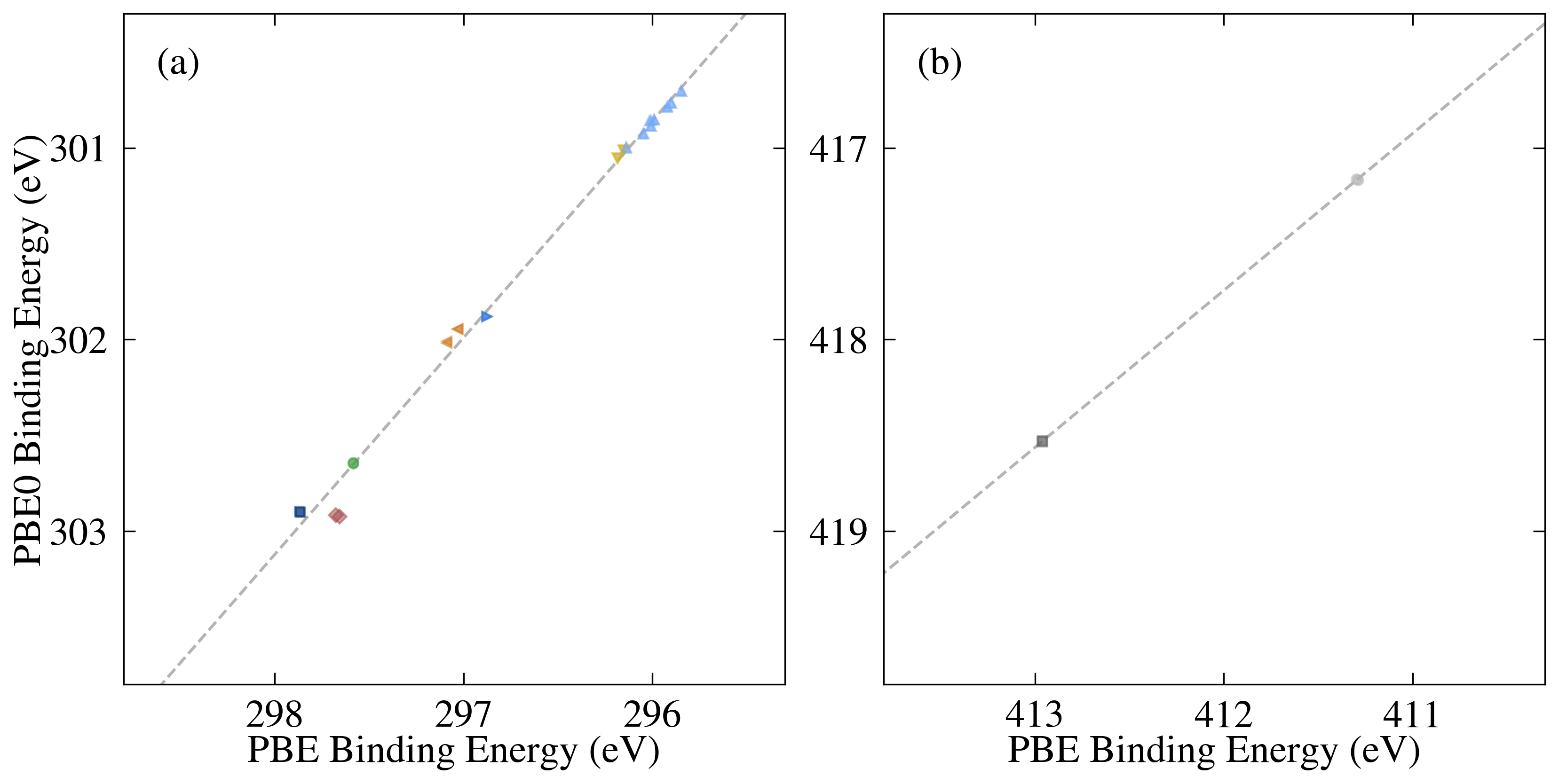}
\caption{Comparison between PBE and PBE0 $\Delta$SCF-calculated core binding energies of the relaxed molecule for (a) C~$1s$ and (b) N~$1s$, where the colours refer to the different atomic sites (see main paper). \label{fig:core_pbe0}}
\end{figure}

\begin{figure}[h]
\centering
\includegraphics[scale=0.37]{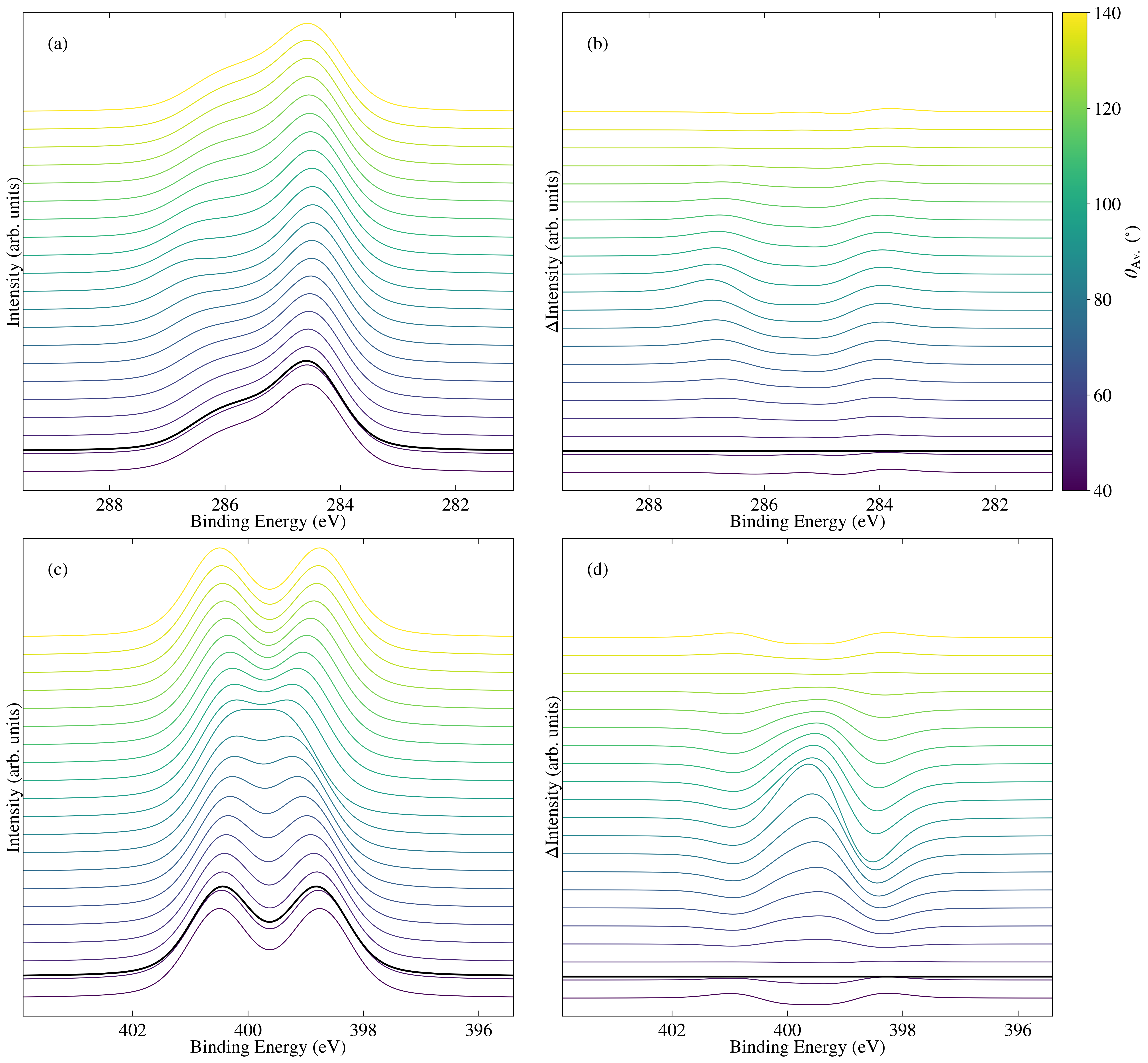}
\caption{Variation in the calculated core spectra with average torsion angle, for the explicitly rotated molecules for (a) C~$1s$ and (c) N~$1s$, where the relaxed molecule is highlighted with thick black lines. Also shown is the difference in density of states with respect to the relaxed molecule for (b) C~$1s$ and (d) N~$1s$. The spectra have been shifted following the alignment described in the main paper. \label{fig:core_theta}}
\end{figure}

\begin{figure}[!ht]
\centering
\includegraphics[scale=0.37]{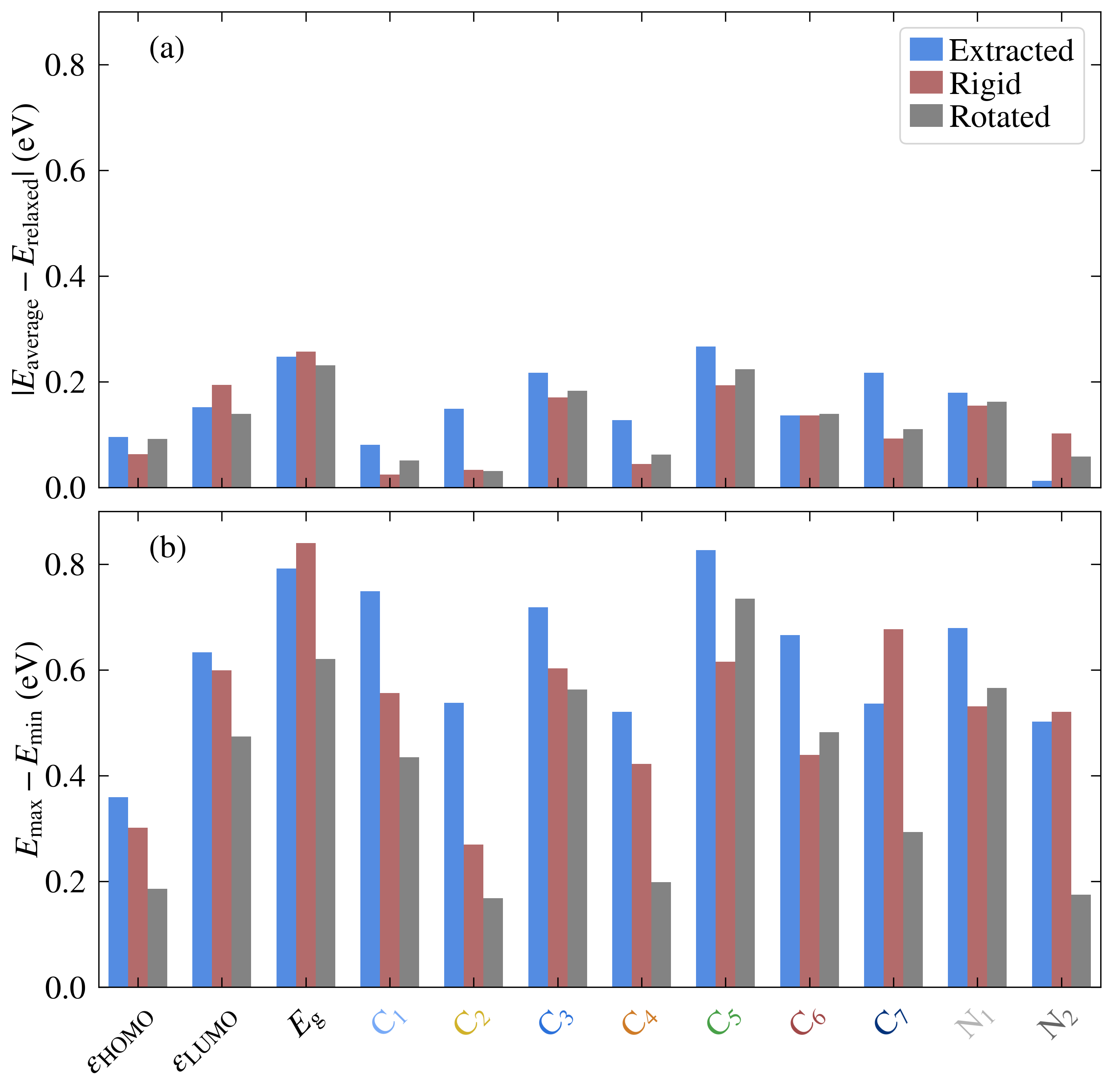}
\caption{Variation in frontier orbital energies and core binding energies for the unmodified extracted, the rigidified and the explicitly rotated molecules. Shown is (a) the difference between the average values and the corresponding value for the relaxed molecule and (b) the difference between the maximum and minimum values for each set of molecules. \label{fig:deviations}}
\end{figure}

\clearpage

\begin{figure}[ht!]
\centering
\includegraphics[scale=0.37]{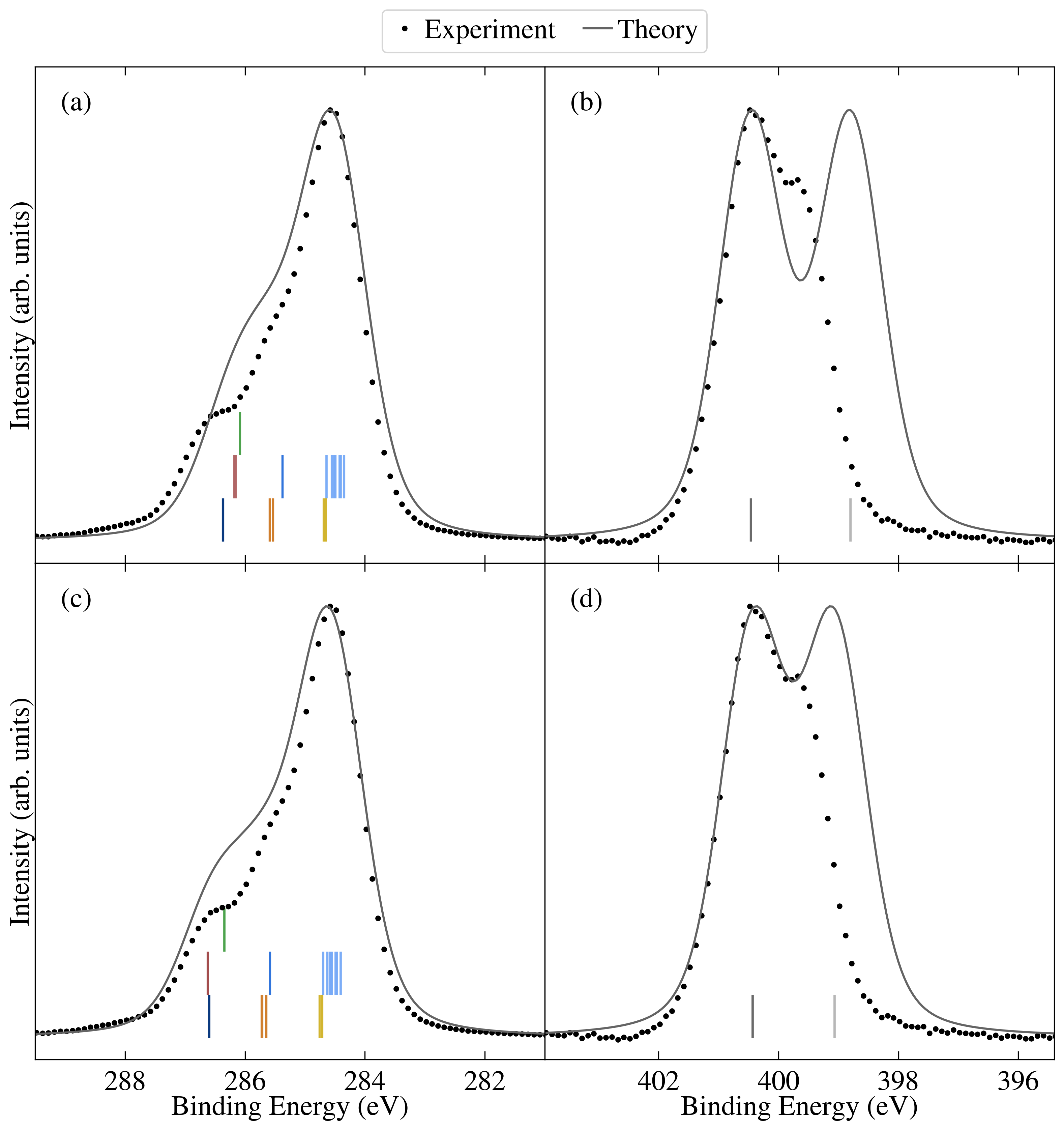}
\caption{Core spectra for the relaxed molecule calculated using PBE for (a) C~$1s$ and (b) N~$1s$, as well as with PBE0 for (c) C~$1s$ and (d) N~$1s$ where the colours refer to the different atomic sites (see main paper). \label{fig:core_xc}}
\end{figure}

\begin{figure}[ht!]
\centering
\includegraphics[scale=0.37]{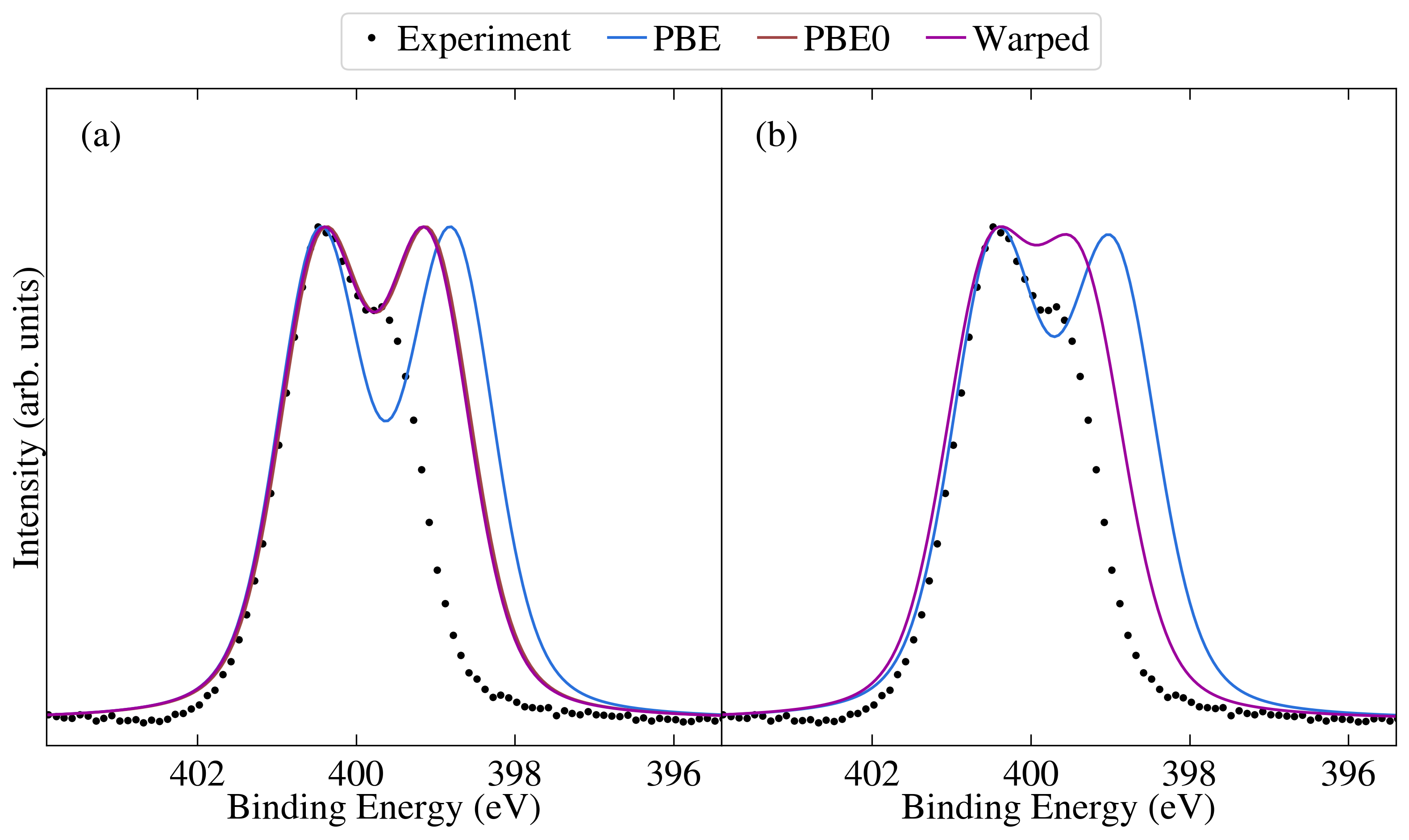}
\caption{Core N~$1s$ spectra calculated using PBE, PBE0, and applying a simple spectral warping for (a) the relaxed molecule, and (b) the average over unmodified extracted molecules. \label{fig:core_warped}}
\end{figure}

\clearpage

\begin{table}[h]
\small
\caption{Summary of calculated values for the frontier orbitals and core binding energies. Average, minimum and maximum values are given across the datasets; in the case where only one value exists (e.g.\ frontier orbitals for the relaxed molecule), minimum and maximum values are omitted.  Values are across all data, i.e.\ for frontier orbitals with PBE this includes all 500 extracted/rigid molecules, while for PBE0 frontier orbitals and PBE core binding energies, values are across the 42 selected molecules only.}
\label{table:data}
\begin{tabular*}{1.0\textwidth}{@{\extracolsep{\fill}}l rrrr rrrr rrrr rrrr}
\hline
 & \multicolumn{4}{c}{Relaxed} &   \multicolumn{4}{c}{Rotated}   &  \multicolumn{4}{c}{Rigid} & \multicolumn{4}{c}{Extracted}  \\
 
\cline{2-5}\cline{6-9}\cline{10-13}\cline{14-17}\\[-2.5ex]
& Av. &  Min. & Max. & Range & Av. &  Min. &  Max. & Range & Av. &  Min. & Max. & Range & Av. &  Min. &  Max. & Range \\
\cline{1-1}\cline{2-2}\cline{3-3}\cline{4-4}\cline{5-5}\cline{6-6}\cline{7-7}\cline{8-8}\cline{9-9}\cline{10-10}\cline{11-11}\cline{12-12}\cline{13-13}\cline{14-14}\cline{15-15}\cline{16-16}\cline{17-17}\\[-2.5ex]

 \textbf{PBE}\\
$\varepsilon_{\mathrm{HOMO}}$  & -5.60 &      - &      - &     - &   -5.50 &  -5.60 &  -5.41 &  0.19 &  -5.53 &  -5.65 &  -5.35 &  0.30 &     -5.53 &  -5.64 &  -5.28 &  0.36 \\
$\varepsilon_{\mathrm{LUMO}}$  & -3.46 &      - &      - &     - &   -3.60 &  -3.82 &  -3.35 &  0.47 &  -3.66 &  -3.96 &  -3.36 &  0.60 &     -3.66 &  -3.97 &  -3.34 &  0.63 \\
$E_{\mathrm{g}}$  &  2.13 &      - &      - &     - &    1.90 &   1.59 &   2.21 &  0.62 &   1.87 &   1.39 &   2.23 &  0.84 &      1.87 &   1.45 &   2.24 &  0.79 \\
C$_1$ &  296.0 &  295.8 &  296.1 &   0.3 &   295.9 &  295.7 &  296.2 &   0.4 &  296.0 &  295.7 &  296.2 &   0.6 &     295.9 &  295.6 &  296.3 &   0.7 \\
C$_2$ &  296.2 &  296.2 &  296.2 &   0.0 &   296.1 &  296.1 &  296.3 &   0.2 &  296.1 &  296.0 &  296.3 &   0.3 &     296.0 &  295.8 &  296.3 &   0.5 \\
C$_3$ &  296.9 &  296.9 &  296.9 &   0.0 &   297.1 &  296.8 &  297.4 &   0.6 &  297.0 &  296.7 &  297.3 &   0.6 &     297.1 &  296.7 &  297.5 &   0.7 \\
C$_4$ & 297.1 &  297.0 &  297.1 &   0.1 &   297.0 &  296.9 &  297.1 &   0.2 &  297.0 &  296.8 &  297.2 &   0.4 &     296.9 &  296.7 &  297.2 &   0.5 \\
C$_5$ &  297.6 &  297.6 &  297.6 &   0.0 &   297.8 &  297.5 &  298.3 &   0.7 &  297.8 &  297.5 &  298.1 &   0.6 &     297.9 &  297.5 &  298.3 &   0.8 \\
C$_6$ &  297.7 &  297.7 &  297.7 &   0.0 &   297.8 &  297.6 &  298.1 &   0.5 &  297.8 &  297.6 &  298.0 &   0.4 &     297.8 &  297.5 &  298.2 &   0.7 \\
C$_7$ & 297.9 &  297.9 &  297.9 &   0.0 &   298.0 &  297.9 &  298.2 &   0.3 &  298.0 &  297.6 &  298.2 &   0.7 &     298.1 &  297.8 &  298.4 &   0.5 \\
N$_1$ & 411.3 &  411.3 &  411.3 &   0.0 &   411.5 &  411.2 &  411.8 &   0.6 &  411.4 &  411.2 &  411.7 &   0.5 &     411.5 &  411.2 &  411.9 &   0.7 \\
N$_2$ & 413.0 &  413.0 &  413.0 &   0.0 &   412.9 &  412.8 &  413.0 &   0.2 &  412.9 &  412.6 &  413.1 &   0.5 &     412.9 &  412.7 &  413.2 &   0.5 \\
 \cline{1-1}\cline{2-2}\cline{3-3}\cline{4-4}\cline{5-5}\cline{6-6}\cline{7-7}\cline{8-8}\cline{9-9}\cline{10-10}\cline{11-11}\cline{12-12}\cline{13-13}\cline{14-14}\cline{15-15}\cline{16-16}\cline{17-17}\\[-2.5ex]
 \textbf{PBE0}\\
$\varepsilon_{\mathrm{HOMO}}$  &   -6.43 &      - &      - &     - &   -6.34 &  -6.42 &  -6.29 &  0.13 &  -6.39 &  -6.48 &  -6.25 &  0.23 &     -6.39 &  -6.53 &  -6.15 &  0.38 \\
$\varepsilon_{\mathrm{LUMO}}$  &   -2.71 &      - &      - &     - &   -2.83 &  -3.00 &  -2.62 &  0.38 &  -2.88 &  -3.10 &  -2.67 &  0.43 &     -2.88 &  -3.10 &  -2.58 &  0.52 \\
$E_{\mathrm{g}}$ &    3.72 &      - &      - &     - &    3.52 &   3.29 &   3.75 &  0.46 &   3.51 &   3.16 &   3.76 &  0.60 &      3.51 &   3.13 &   3.74 &  0.61 \\
C$_1$ &  300.8 &  300.7 &  301.0 &   0.3 &       - &      - &      - &     - &      - &      - &      - &     - &         - &      - &      - &     - \\
C$_2$ &  301.0 &  301.0 &  301.1 &   0.0 &       - &      - &      - &     - &      - &      - &      - &     - &         - &      - &      - &     - \\
C$_3$ & 301.9 &  301.9 &  301.9 &   0.0 &       - &      - &      - &     - &      - &      - &      - &     - &         - &      - &      - &     - \\
C$_4$ &  302.0 &  301.9 &  302.0 &   0.1 &       - &      - &      - &     - &      - &      - &      - &     - &         - &      - &      - &     - \\
C$_5$ & 302.6 &  302.6 &  302.6 &   0.0 &       - &      - &      - &     - &      - &      - &      - &     - &         - &      - &      - &     - \\
C$_6$ &   302.9 &  302.9 &  302.9 &   0.0 &       - &      - &      - &     - &      - &      - &      - &     - &         - &      - &      - &     - \\
C$_7$ &  302.9 &  302.9 &  302.9 &   0.0 &       - &      - &      - &     - &      - &      - &      - &     - &         - &      - &      - &     - \\
N$_1$ & 417.2 &  417.2 &  417.2 &   0.0 &       - &      - &      - &     - &      - &      - &      - &     - &         - &      - &      - &     - \\
N$_2$ &  418.5 &  418.5 &  418.5 &   0.0 &       - &      - &      - &     - &      - &      - &      - &     - &         - &      - &      - &     - \\
  
\hline
\end{tabular*}
\end{table}
